\documentclass[aps,pra,twocolumn,amsmath,amssymb,showpacs]{revtex4-1}

\newcommand{\Jnature}{Nature (London)}
\newcommand{\Jnatphys}{Nat. Phys.}

\newcommand{\Jscience}{Science}

\newcommand{\Jprl}{Phys. Rev. Lett.}

\newcommand{\Jpra}{Phys. Rev. A}
\newcommand{\Jprb}{Phys. Rev. B}

\newcommand{\Jpre}{Phys. Rev. E}

\newcommand{\Jepl}{Europhys. Lett.}

\newcommand{\Jepjd}{Eur. Phys. J. D}

\newcommand{\Jjetp}{Sov. Phys. JETP}
\newcommand{\JjphysUSSR}{J. Phys. USSR}
\newcommand{\JSovJlowT}{Sov. J. Low Temp. Phys.}

\newcommand{\JTheorMathPhys}{Theor. Math. Phys.}

\newcommand{\JjphysquatreF}{J. Phys. IV (France)}

\usepackage{graphicx}
\usepackage{dcolumn}
\usepackage{bm}
\usepackage{color}
\usepackage{hyperref}
\hypersetup{
colorlinks=true,
citecolor=blue,
linkcolor=red,
urlcolor=black
}

\newcommand{\kB}{k_{\textrm{\tiny B}}}

\def\Li6{$^6$Li}
\def\K40{$^{40}$K}

\def\vk{\mathbf{k}}

\newcommand{\la}{\langle}
\newcommand{\ra}{\rangle}
\newcommand{\psih}{\hat{\psi}}
\newcommand{\psihd}{\hat{\psi}^\dagger}
\newcommand{\bh}{\hat{b}}
\newcommand{\bhd}{\hat{b}^\dagger}

\newcommand{\Lh}{\hat{\Lambda}}

\newcommand{\te}{\theta}
\newcommand{\teh}{\hat{\theta}}

\newcommand{\n}{n}
\newcommand{\nh}{\hat{n}}

\newcommand{\nt}{\hat{X}}

\newcommand{\tet}{\hat{P}}

\newcommand{\rr}{\mathbf{r}}
\newcommand{\dd}{\textrm{d}}

\newcommand{\ham}{\hat{\mathcal{H}}}
\newcommand{\hamGC}{\hat{H}}

\newcommand{\LGP}{\mathcal{L}^{\textrm{\tiny GP}}}
\newcommand{\bigB}{\mathcal{B}}
\newcommand{\bigBbar}{\bar{\mathcal{B}}}

\newcommand{\Eof}{E_{\vk}^{\textrm{off}}}
\newcommand{\Ein}{E_{\vk}^{\textrm{in}}}
\newcommand{\inb}{"in" branch }
\newcommand{\offb}{"off" branch }

\begin{document}

\title{Two-component Bose gases with one-body and two-body couplings}
\date{\today}
\author{Samuel Lellouch}
\author{Tung-Lam Dao}
\author{Thomas Koffel}
\author{Laurent Sanchez-Palencia}
\affiliation{
Laboratoire Charles Fabry,
Institut d'Optique, CNRS, Univ Paris Sud,
2 avenue Augustin Fresnel,
F-91127 Palaiseau cedex, France}

\begin{abstract}
We study the competition between one-body and two-body couplings in weakly-interacting two-component Bose gases, in particular as regards field correlations.
We derive the meanfield theory for both ground state and low-energy pair excitations in the general case where both one-body and two-body couplings are position-dependent and the fluid is subjected to a state-dependent trapping potential. General formulas for phase and density correlations are also derived.
Focusing on the case of homogeneous systems, we discuss the pair-excitation spectrum and the corresponding excitation modes, and use them to calculate correlation functions, including both quantum and thermal fluctuation terms.
We show that the relative phase of the two components is imposed by that of the one-body coupling, while its fluctuations are determined by the modulus of the one-body coupling and by the two-body coupling. One-body coupling and repulsive two-body coupling cooperate to suppress relative-phase fluctuations, while attractive two-body coupling tends to enhance them.
Further applications of the formalism presented here and extensions of our work are also discussed.
\end{abstract}


\maketitle

\section{Introduction}
Multi-component (spinor) quantum fluids underlie a variety of physical systems, such as $^3$He-$^4$He mixtures in three-fluid models~\cite{andreev1975}, Bose-condensed spin-polarized hydrogen gases in the two lowest-energy states~\cite{mullin1980,siggia1980,statt1980}, optically-excited excitons in high-quality Cu$_2$0 crystals~\cite{snoke1990,lin1993}, as well as gaseous Bose-Einstein condensates either in two overlapped atomic hyperfine states~\cite{matthews1998,hall1998a,hall1998b} or in adjacent traps coupled by tunnel effect~\cite{schumm2005b}. The dynamics of spinors sparks a variety of physical effects, including quantum phase transitions, topological defects, and spin domains, governed by the complex interplay of particle-particle interaction, exchange coupling, magnetic-like ordering, and temperature effects.
Early studies focused on the possibility of observing Bose-Einstein condensation~\cite{shi1995}, as well as stability conditions~\cite{andreev1975,goldstein1997,ohberg1999}, phase separation~\cite{ho1996,law1997,bashkin1997,busch1997,hall1998a,timmermans1998,pu1998a,pu1998b}, and spontaneous symmetry breaking mechanisms~\cite{ohberg1998,gordon1998,esry1999,sadler2006}
in two-component Bose-Einstein condensates.
Two-component Bose gases have also been used to study phase coherence~\cite{javanainen1996},
Josephson like physics~\cite{whitlock2003,chang2005,stimming2010,zibold2010,betz2011},
the dynamics of spin textures~\cite{kamatsu2005,vengalattore2008,vengalattore2010,guzman2011},
random-field-induced order effects~\cite{wehr2006,niederberger2008},
and twin quantum states for quantum information processing~\cite{cirac1998,gross2011,lucke2011}.

In the context of ultracold gases the combination of optical and magnetic fields designed to manipulate the internal states of alkali-metal atoms offer a wide range of possibilities to accurately engineer multi-component quantum fluids. Such systems offer a new tool to study quantum coherence in various contexts~\cite{javanainen1996,hall1998b,chang2005,betz2011}. For instance, measurement of the relative-phase correlation function of a coupled binary Bose gas in one dimension was reported in Ref.~\cite{betz2011}. In the latter case, the coupling was of the Josephson (one-body) type.

\begin{figure}[t!]
\vspace{0.1725cm}
\includegraphics[width=8.7cm,clip=true]{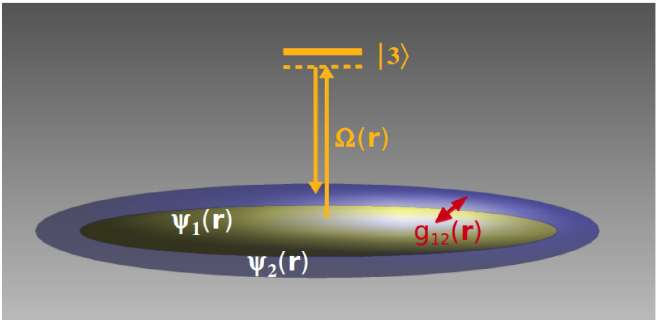}
\caption{\label{fig:coupledBECs}
Coupled two-component Bose gas. The gas is made of bosonic particles of a single atomic species,
which can be in two different internal states (labeled $1$ and $2$).
It is described by the two field operators $\hat{\psi}_1(\rr)$ and $\hat{\psi}_2(\rr)$, corresponding to each component. In this work, we assume that the two components are coupled by one-body and/or two-body interactions of coupling constants $\Omega$ and $g_{12}$, respectively. In the most general case, the two coupling constants can be position dependent.
}
\end{figure}

In this paper, we consider a two-component Bose gas with both one-body (field-field) and two-body (density-density) couplings, and focus our analysis on the pair-excitation spectrum and the relative phase correlation function at both zero and finite temperature. The most general case can be realized in ultracold-atom gases by using a mixture of atoms in two different internal hyperfine states (noted $1$ and $2$) of the same atomic species. The two-body interaction with coupling constant $g_{12}$ results from short-range particle-particle interactions between atoms in different internal states, while the one-body interaction can be implemented by two-photon Raman optical coupling, which transfers atoms from one internal state to the other (see schematic view on Fig.~\ref{fig:coupledBECs}).
In Sec.~\ref{sec:MFtheory}, we present the model and derive the meanfield theory of the coupled two-component Bose fluid for both ground state and low-energy pair excitations. The theory is formulated in the most general case, where both one-body and two-body couplings are position dependent and the fluid is subjected to a state-dependent trapping potential. In addition, we use the phase-density Bogoliubov-Popov approach, which allows us to treat true condensates and quasi-condensates on equal footing~\cite{popov1972,popov1983}. General formulas for phase and density correlations are derived.
In Sec.~\ref{sec:homog}, we focus on the case of homogeneous systems, which allow considerable simplification of the formalism and contain most of the physical effects. After rewriting the general meanfield equations for homogeneous systems~(Sec.~\ref{sec:homog.rewrite}), we discuss the pair-excitation spectrum and the corresponding fields, and use them to calculate the correlation functions including both quantum and thermal fluctuation terms~(Sec.~\ref{sec:homog.general}).
Our main conclusions are as follows:
The phase of the one-body coupling term imposes alone the relative phase of the two components at the meanfield background level. Then, the fluctuations of the relative phase are determined by the interplay of the modulus of the one-body term and the two-body term. On the one hand, the one-body coupling always favors local mutual coherence of the two components but the correlation length decreases when the modulus of the one-body term increases. On the other hand, repulsive two-body coupling cooperates with one-body coupling to further suppress relative-phase fluctuations, while attractive two-body coupling competes with one-body coupling to enhance relative-phase fluctuations.
These results are summarized in more detail in Sec.~\ref{sec:conclusion}, where we also discuss further possible applications of the formalism presented here.

\section{Mean-field theory of a two-component Bose gas}
\label{sec:MFtheory}
Consider a two-component Bose-Bose mixture at thermodynamic equilibrium at temperature $T$,
and in the weakly interacting regime.
We assume that the two components (labelled by $\sigma\in\{1,2\}$) interact with each other
and can exchange atoms to maintain chemical equilibrium. 
The average total number of atoms, $N=N_1+N_2$, is conserved but
the average number of atoms in each component, $N_{\sigma}$, is not.
The physics of this system is governed by the grand-canonical Hamiltonian
\begin{equation}
\hamGC \equiv \ham-\mu\hat{N} = \hamGC_1 + \hamGC_2 + \hamGC_{12} \,,
\label{eq:K}
\end{equation}
where $\ham$ is the many-body Hamiltonian and
$\hat{N} = \hat{N}_1 + \hat{N}_2$ is the total number operator,
with $\hat{N}_\sigma = \int\dd\rr\ \psihd_\sigma(\rr)\psih_\sigma(\rr)$
and
$\psih_{\sigma}(\rr)$ the (bosonic) field operator of component $\sigma$.
Assuming two-body contact interactions,
the Hamiltonian associated the the sole component $\sigma$
(written in the grand-canonical form for the chemical potential $\mu$ of the mixture) is
\begin{equation}
\hamGC_\sigma =
\int \dd\rr\ \psihd_{\sigma} \left[
-\frac{\hbar^2\nabla^2}{2m}
+V_\sigma-\mu
+\frac{g_\sigma (\rr)}{2} \psihd_{\sigma}\psih_{\sigma}
\right] \psih_{\sigma}
\label{eq:Ksigma}
\end{equation}
and the coupling Hamiltonian is
\begin{equation}
\hamGC_{12}=\int\!\!\dd\rr \left[\ g_{12}(\rr)\psihd_{1}\psihd_{2}\psih_{1}\psih_{2} +
\left(\frac{\hbar\Omega (\rr)}{2}\psihd_{2}\psih_{1}+\textrm{H.c.}\right) \right]\,.
\label{eq:K12}
\end{equation}
The single-component Hamiltonian $\hamGC_\sigma$ contains
(i)~a kinetic term ($m$ is the atomic mass),
(ii)~a potential term, $V_\sigma (\rr)$, both associated with single-particle dynamics,
and
(iii)~an intra-component interaction term of coupling parameter $g_\sigma$.
The coupling Hamiltonian, $\hamGC_{12}$, contains
(i)~a term originating from elastic contact interaction between two atoms in different components
characterized by the inter-component coupling constant $g_{12}$,
and
(ii)~an exchange term proportional to $\Omega$, which transfers atoms from one component to the other
and in particular permits chemical equilibrium.
In ultracold-atom systems,
the exchange one-body term can be realized by two-photon Raman or radio-frequency coupling~\cite{matthews1998}
or by Josephson coupling between two adjacent traps~\cite{raghavan1999,whitlock2003,albiez2005,gati2006,betz2011}, whereas the two-body coupling can be controlled by Feshbach resonance techniques~\cite{inouye1998}.
In the most general case, all coupling terms $g_1$, $g_2$, $g_{12}$, and $\Omega$ can be position-dependent. Hereafter, we write $\Omega (\rr) \equiv \Omega_0 (\rr) \textrm{e}^{-i\alpha(\rr)}$, with $\Omega_0 = |\Omega|$ and $\alpha(\rr)$ the phase of the exchange coupling, for convenience.

In the following,
we first reformulate the above Hamiltonians into the phase-density formalism, which is more appropriate for our study.
We then apply the Gross-Pitaevskii approach, which describes the meanfield quasicondensate background of the two-component Bose-Bose mixture,
and develop the Bogoliubov-de Gennes theory for the mixture, which provides the spectrum of collective excitations
and can be used to describe finite-temperature effects.
We finally write the general expressions for the density and phase correlation functions,
which are calculated in the next sections.
Although the process we follow is standard, we generalize previous work to the case where their
couplings can be position-dependent. We thus detail the derivation of the main equations.

\subsection{Phase-density formalism}
\label{sec:MFtheory.PhaseDens}
The complete grand-canonical Hamiltonian~$\hamGC$ is invariant under the gauge transformation
$\{\psih_1 (\rr),\psih_2 (\rr)\} \rightarrow \textrm{e}^{i\theta_0}\{\psih_1 (\rr),\psih_2 (\rr)\}$
for any value of $\theta_0 \in \mathbb{R}$, as can be easily checked in Eqs.~(\ref{eq:Ksigma}) and (\ref{eq:K12}).
More precisely, if $\Omega (\rr) \equiv 0$, the phases of the two components are independent and
$\hamGC$ is invariant under the more general transformation
$\{\psih_1 (\rr),\psih_2 (\rr)\} \rightarrow \{\textrm{e}^{i{\theta_0^1}}\psih_1 (\rr),\textrm{e}^{i{\theta_0^2}}\psih_2 (\rr)\}$
for any values of $\theta_0^1,\theta_0^2 \in \mathbb{R}$.
If however $\Omega (\rr) \not\equiv 0$, the phases of the two components are coupled via the last term in Eq.~(\ref{eq:K12})
and the relative phase is a determined quantity.
In both cases, the phases of the field operators $\psih_\sigma (\rr)$ are not fully determined and it is useful to turn to the phase-density formalism.
The latter is successfully used in the literature for a long time~\cite{popov1972,shevchenko1992}
and was recently developed in a lattice formulation, which allows for a precise definition of the phase
operator~\cite{mora2003}.
We write the field operator for each component in the form
\begin{equation}
\psih_\sigma(\rr)=\textrm{e}^{i\teh_\sigma(\rr)}\sqrt{\nh_\sigma(\rr)},
\label{eq:phase-density}
\end{equation}
where the density ($\nh_\sigma$) and phase ($\teh_\sigma$) operators satisfy the Bose commutation rule
$[\nh_{\sigma}(\rr),\teh_{\sigma'}(\rr')]=i\delta_{\sigma\sigma'}\delta(\rr-\rr')$.
Replacing $\psih_\sigma$ by expression~(\ref{eq:phase-density}) into Eqs.~(\ref{eq:Ksigma}) and (\ref{eq:K12}),
we find
\begin{equation}
\hamGC_{\sigma} =
\int \!\! \dd\rr\ \sqrt{\nh}_{\sigma} \Big[
  \frac{-\hbar^2}{2m} \left( \nabla^2 \! - \! |\nabla\teh_{\sigma}|^2 \right)
  + V_{\sigma}-\mu
  +\frac{g_\sigma}{2}\nh_{\sigma}
\Big] \sqrt{\nh}_{\sigma}
\label{eq:KsigmaBIS}
\end{equation}
and
\begin{equation}
\hamGC_{12}
=
\int \!\! \dd\rr\ \Big[
   g_{12}\nh_{1}\nh_{2}
  + \left\{ \frac{\hbar\Omega}{2} \sqrt{\nh_2} \textrm{e}^{i(\teh_1 - \teh_2)} \sqrt{\nh_1} + \textrm{H.c.}
    \right\}
\Big] \,.
\label{eq:K12BIS}
\end{equation}
Expressions~(\ref{eq:KsigmaBIS}) and (\ref{eq:K12BIS}) determine the complete Hamiltonian~(\ref{eq:K}) in terms of density and phase operators~\footnote{Note that we have dropped a constant term arising from the commutation relation of $\psih_\sigma (\rr)$ and $\psihd_\sigma (\rr)$ in the intra-component interaction term of Eq.~(\ref{eq:Ksigma}). The latter can be absorbed in a renormalization of the chemical potential $\mu$.}.
This form is particularly suitable for perturbative expansion in the condensate or quasi-condensate regime, where the density fluctuations are suppressed by strong-enough repulsive interactions but the phase fluctuations can be large~\cite{popov1972,popov1983,petrov2000b,mora2003,petrov2004}.

\subsection{Meanfield background: Gross-Pitaevskii theory}
The zeroth-order term in quantum and thermal fluctuations corresponds to the meanfield background. The latter is determined using the Gross-Pitaevskii approach~\cite{pethick2001,pitaevskii2004}, adapted to the two-component mixture. It amounts to minimize the grand-canonical energy functional
$E_{\textrm{\tiny MF}} \equiv \langle \psi_{\textrm{\tiny MF}} \vert \hamGC \vert \psi_{\textrm{\tiny MF}} \rangle$ with the
two-component Hartree-Fock ansatz
\begin{equation}
\vert \psi_{\textrm{\tiny MF}} \rangle =
  \frac{(\hat{a}_1^\dagger)^{N_1}}{\sqrt{N_1 !}} \
  \frac{(\hat{a}_2^\dagger)^{N_2}}{\sqrt{N_2 !}} \
 \vert \textrm{vac} \rangle \,,
\label{eq:MFansatz}
\end{equation}
where $\hat{a}_\sigma^\dagger$ creates an atom in component $\sigma$ with a spatial
wave function $\psi_\sigma (\rr) \equiv \textrm{e}^{i\te_\sigma(\rr)}\sqrt{\n_\sigma(\rr)}$.
At this stage, the number of atoms in each component, $N_\sigma$, and the corresponding
phase [$\te_\sigma(\rr)$] and density [$\n_\sigma(\rr)$] fields
are unknown variational quantities.
Here, we use the normalization condition $\int \dd\rr\ \n_\sigma (\rr) = N_\sigma$
and we recall that the chemical potential $\mu$ is determined implicitly by the relation
$\int \dd\rr\ [\n_1 (\rr) + \n_2 (\rr) ] = N$.

Proceeding in the standard way, we evaluate the complete grand-canonical Hamiltonian~(\ref{eq:K})
within the Hartree-Fock ansatz~(\ref{eq:MFansatz}) and find
\begin{equation}
E_\textrm{\tiny MF}
=
\langle \hamGC_1 \rangle_\textrm{\tiny MF}
+
\langle \hamGC_2 \rangle_\textrm{\tiny MF}
+
\langle \hamGC_{12} \rangle_\textrm{\tiny MF}
\label{eq:Emf}
\end{equation}
where $\langle \hamGC_\sigma \rangle_\textrm{\tiny MF}$ and $\langle \hamGC_{12} \rangle_\textrm{\tiny MF}$
are given by Eqs.~(\ref{eq:KsigmaBIS}) and (\ref{eq:K12BIS}) with
the phase $\teh_\sigma (\rr)$ and density $\nh_\sigma (\rr)$ operators
replaced by the corresponding Hartree-Fock fields $\te_\sigma (\rr)$ and $\n_\sigma (\rr)$.
Then, minimizing $E_{\textrm{\tiny MF}}$ with respect to $\te_\sigma (\rr)$ and $\n_\sigma (\rr)$
yields the following coupled Euler-Lagrange equations:

\begin{eqnarray}
0 & = &
-\frac{\hbar^2}{2m} \left( \frac{\nabla^2 \sqrt{n_\sigma}}{\sqrt{n_\sigma}} - |\nabla\te_{\sigma}|^2 \right)
  + V_{\sigma}-\mu
  + g_\sigma \n_\sigma + g_{12}\n_{\bar{\sigma}}
  \nonumber \\
&&
+ \frac{\hbar\Omega_0}{2} \sqrt{\frac{n_{\bar{\sigma}}}{n_\sigma}} \cos (\te - \alpha)
\label{eq:GPE1} \\
0 & = &
\frac{\hbar^2}{m}\nabla(\n_\sigma\nabla\te_\sigma) \pm \hbar\Omega_0\sqrt{n_1n_2}\sin (\theta - \alpha) \,,
\label{eq:GPE2}
\end{eqnarray}
where
$\te (\rr) \equiv \te_1 (\rr) - \te_2 (\rr)$ is the relative phase between the two components,
$\bar{\sigma}$ is the conjugate of $\sigma$ [i.e.\ $\bar{\sigma}=2$ (resp.\ $1$) for $\sigma = 1$ (resp.\ $2$)],
and the $\pm$ sign in Eq.~(\ref{eq:GPE2}) is $+$ (resp.\ $-$) for $\sigma=1$ (resp.\ $2$).

\subsection{Excitations: Bogoliubov-de Gennes theory}
The low-energy spectrum of the collective excitations of the two-component Bose gas is then determined
using the Bogoliubov-de~Gennes approach~\cite{bogoliubov1947,bogoliubov1958,degennes1995,popov1972,popov1983},
which amounts to perform a perturbative expansion of Hamiltonian~(\ref{eq:K})
in phase and density fluctuations. We write
$\nh_\sigma = \n_\sigma + \delta\nh_\sigma$
and $\teh_\sigma =  \te_\sigma + \delta\teh_\sigma$,
with $\n_\sigma(\rr)$ and $\te_\sigma(\rr)$ given by the mean-field Gross-Pitaevskii theory,
and
\begin{equation}
\vert \delta\nh_\sigma \vert \ll n_\sigma
\quad \textrm{and} \quad
\vert\nabla\delta \teh_{\sigma}\vert \ll m c / \hbar
\end{equation}
where $c=\sqrt{\mu/m}$ is the velocity of sound in a single-component Bose-Einstein (quasi-)condensate of chemical potential $\mu$.
These conditions are usually well verified in weakly-interacting ultracold, two-component gases~\cite{myatt1997,matthews1998,hall1998a,hall1998b}.

\subsubsection{Weak-fluctuation expansion of the Hamiltonian}
Proceeding up to second order in phase and density fluctuations,
it is convenient to define the position-dependent operators
\begin{equation}
\nt_\sigma (\rr) \equiv \frac{\delta\nh_\sigma (\rr)}{2\sqrt{n_\sigma (\rr)}}
\label{eq:X}
\end{equation}
and
\begin{equation}
\tet_\sigma (\rr) \equiv \sqrt{n_\sigma (\rr)}\delta\teh_\sigma (\rr)\,,
\label{eq:P}
\end{equation}
which are canonical conjugates (up to a multiplying factor of $1/2$),
i.e.\
$\left[\nt_{\sigma}(\rr),\tet_{\sigma^\prime}(\rr^\prime)\right]=i\delta_{\sigma,\sigma^\prime}\delta\left(\rr-\rr^\prime\right)/2$.
Then, inserting
$\sqrt{\nh_\sigma} \simeq \sqrt{\n_\sigma} + \nt_\sigma - \nt_\sigma^2/2\sqrt{\n_\sigma}$
and $\teh_\sigma = \te_\sigma + \tet/\sqrt{\n_\sigma}$
into Eqs.~(\ref{eq:KsigmaBIS}) and (\ref{eq:K12BIS}),
we find
\begin{equation}
\hamGC \simeq E_{\textrm{\tiny MF}} + \hamGC_1^{(2)} + \hamGC_2^{(2)} + \hamGC_{12}^{(2)} \,.
\end{equation}
The zeroth-order term, $E_{\textrm{\tiny MF}}$, coincides with the mean-field energy~(\ref{eq:Emf})
where the fields $\n_\sigma$ and $\te_\sigma$ are substituted to the solutions of the coupled
Euler-Lagrange equations~(\ref{eq:GPE1}) and (\ref{eq:GPE2}).
The first-order term,
$\hamGC^{(1)}=\sum_\sigma\left\{\delta\nh_\sigma\cdot\left.\frac{\partial\hamGC}{\partial\nh_\sigma}\right\vert_{\psi_{\textrm{\tiny MF}}} + \delta\teh_\sigma\cdot\left.\frac{\partial\hamGC}{\partial\teh_\sigma}\right\vert_{\psi_{\textrm{\tiny MF}}} \right\}$,
vanishes since the zeroth-order term minimizes
$\langle \psi_{\textrm{\tiny MF}} \vert \hamGC \vert \psi_{\textrm{\tiny MF}} \rangle = E_{\textrm{\tiny MF}}$.
The second-order terms,
$\hamGC_1^{(2)}, \hamGC_2^{(2)}$ and $\hamGC_{12}^{(2)}$,
are found after some straightforward algebra, which yields
\begin{eqnarray}
\hat{H}^{(2)}_\sigma
&=&
\int \dd\rr\
  \nt_\sigma \left[-\frac{\hbar^2}{2m} \left(\nabla^2-\frac{\nabla^2\sqrt{n_\sigma}}{\sqrt{n_\sigma}}\right)+2g_\sigma n_\sigma\right] \nt_\sigma
\nonumber \\
&&
+ \int \dd\rr\
  \tet_\sigma\left[-\frac{\hbar^2}{2m}\left(\nabla^2-\frac{\nabla^2\sqrt{n_\sigma}}{\sqrt{n_\sigma}}\right)\right]\tet_\sigma
\label{eq:H2sigma} \\
&&
+ \int \dd\rr\
  \frac{2\hbar^2}{m} \nabla\theta_\sigma\cdot \left(\sqrt{n_\sigma} \nt_\sigma\right) \nabla \left(  \tet_\sigma / \sqrt{n_\sigma} \right)\,,
\nonumber 
\end{eqnarray}
where some irrelevant constant terms have been dropped,
and
\begin{eqnarray}
\hat{H}_{12}^{(2)}
&=&
-\sum_\sigma \int\dd\rr\ \frac{\hbar\Omega_0}{2}\sqrt{\frac{n_{\bar{\sigma}}}{n_{\sigma}}}\cos(\theta-\alpha)
\left[\nt_\sigma^2 + \tet_\sigma^2\right]  \nonumber \\
&&
+ \int\!\!\dd\rr\ \Big[4g_{12}\sqrt{n_1n_2}+\hbar\Omega_0\cos(\te-\alpha)\Big]\nt_1\nt_2 
\nonumber \\
&&
+ \int\!\!\dd\rr\ \hbar\Omega_0\cos(\te-\alpha)\tet_1\tet_2
\label{eq:H212} \\
&&
+\int\!\!\dd\rr\ \hbar\Omega_0\sin(\te-\alpha)\Big[\nt_1\tet_2-\nt_2\tet_1 \Big]
\nonumber \\
&&
-\int\!\!\dd\rr\ \hbar\Omega_0\sin(\te-\alpha)\left[\frac{\sqrt{n_2}}{\sqrt{n_1}}\nt_1\tet_1
                                                     -\frac{\sqrt{n_1}}{\sqrt{n_2}}\nt_2\tet_2\right] \,.
\nonumber
\end{eqnarray}

We now apply the canonical transformation~\footnote{This transformation simply arises by analogy with the annihilation operator of the harmonic oscillator. Here, the density fluctuation operator $\delta\nh_\sigma/2\sqrt{\n_\sigma}$ plays the same role as the position operator $\nt_\sigma$ and the phase fluctuation operator $\sqrt{\n_\sigma}\delta\teh$ plays the same role as the momentum operator $\tet_\sigma$ of the quantum harmonic oscillator~\cite{basdevant2005}.} to our quadratic Hamiltonian~\footnote{In the case of a pure condensate with macroscopic occupation of a unique single-particle state, $\psi_\sigma$ (assumed to be real-valued), the operator $\hat{B}_\sigma$ represents the fluctuations of the field operator: $\hat{\psi}_\sigma \simeq \psi_\sigma + \hat{B}_\sigma$.}
\begin{equation}
\hat{B}_\sigma \equiv \nt_\sigma + i\tet_\sigma \,,
\end{equation}
such that the operators $\hat{B}_\sigma$ satisfy the Bose commutation rules
\begin{eqnarray}
&& {[{\hat{B}_{\sigma}}(\rr),{\hat{B}_{\sigma^\prime}}(\rr^\prime)] = 0}
\label{eq:commutBa} \\
&& {[{\hat{B}}_{\sigma}(\rr),{\hat{B}}^{\dagger}_{\sigma^\prime}(\rr^\prime)] = \delta_{\sigma\sigma^\prime}\delta(\rr-\rr^\prime)} \,.
\label{eq:commutBb}
\end{eqnarray}
Then, summing all contributions of Eq.~(\ref{eq:H2sigma}) for $\sigma=1$ and $\sigma=2$
and those of Eq.~(\ref{eq:H212}), we find
\begin{eqnarray}
\hamGC^{(2)}
&=&
\frac{1}{2}\sum_\sigma\int \dd\rr\
  \Big[
    \hat{B}^\dagger_\sigma\textbf{A}_\sigma\hat{B}_\sigma+\hat{B}_\sigma\textbf{A}^*_\sigma\hat{B}^\dagger_\sigma
\nonumber \\
&&    \hspace{2.cm} + \left\{g_\sigma n_\sigma\hat{B}_\sigma\hat{B}_\sigma + \textrm{H.c.}\right\}
  \Big]
\label{eq:H2one} \\
&&
+\int \dd\rr\ \Big[g_{12}\sqrt{\n_1\n_2}\hat{B}_1\hat{B}_2 + \textrm{H.c.} \Big]
\nonumber \\
&& +\int \dd\rr\ \left[\left\{g_{12}\sqrt{\n_1\n_2}+\frac{\hbar\Omega}{2}e^{i\te}\right\}\hat{B}^\dagger_2\hat{B}_1 + \textrm{H.c.} \right]
\nonumber
\end{eqnarray}
where we have used the coupled Euler-Lagrange equation~(\ref{eq:GPE1}) to simplify a couple of terms,
and have introduced the super-operator
\begin{eqnarray}
\textbf{A}_\sigma
&=&
-\frac{\hbar^2}{2m} \left(
    \nabla^2 + 2i \nabla\te_\sigma\cdot\nabla - \vert\nabla\te_\sigma\vert^2\right)
+ V_\sigma-\mu 
\nonumber \\
&& + 2g_\sigma n_\sigma + g_{12}n_{\bar{\sigma}}\,.
\end{eqnarray}
Finally, the Hamiltonian~(\ref{eq:H2one}) can be written in a more compact form by introducing the four-component operators
\begin{equation}
\bigBbar \equiv \left[\hat{B}^\dagger_1, -\hat{B}_1, \hat{B}^\dagger_2, -\hat{B}_2 \right]
\quad \textrm{and} \quad
\bigB \equiv
	\left[
	    \begin{array}{c}
	    \hat{B}_1 \\
	    \hat{B}^\dagger_1 \\
	    \hat{B}_2 \\
	    \hat{B}^\dagger_2
	    \end{array}
	\right]
\end{equation}
so that
\begin{equation}
\hamGC^{(2)}=\frac{1}{2}\int \dd\rr\
  \bigBbar (\rr) \textbf{M} (\rr) \bigB (\rr)
+\textrm{const}
\label{eq:H2two}
\end{equation}
where $\textbf{M} (\rr)$ is the $4 \times 4$ super-operator defined by
\begin{equation}
\textbf{M} \equiv
	\left[
	    \begin{array}{cc}
	    \LGP_1 & \Gamma \\
	    \Gamma^* & \LGP_2\\
	    \end{array}
	\right]
\end{equation}
with
\begin{equation}
\LGP_\sigma=\left[
	\begin{array}{cc}
	    +\textbf{A}_\sigma & +g_\sigma n_\sigma \\
	    -g_\sigma n_\sigma & -\textbf{A}^*_\sigma
	\end{array}
\right]
\end{equation}
and
\begin{equation}
\Gamma=\left[
	\begin{array}{cc}
	    +g_{12}\sqrt{n_1n_2} + \frac{\hbar\Omega^*}{2} e^{-i\theta} & +g_{12}\sqrt{n_1n_2} \\
	    - g_{12}\sqrt{n_1n_2} & - g_{12}\sqrt{n_1n_2} - \frac{\hbar\Omega}{2} e^{+i\theta}
	\end{array}
\right] \,.
\end{equation}

\subsubsection{Bogoliubov transformation}
The second-order term~(\ref{eq:H2two}) in the expansion of the many-body Hamiltonian~(\ref{eq:K}) governs the low-energy excitations of the two-component Bose gas. Its quadratic form is convenient for diagonalization via the usual Bogoliubov method~\cite{bogoliubov1947,bogoliubov1958,popov1972,popov1983}, adapted to the two-component Bose gas. Here, we extend previous approaches~\cite{goldstein1997,whitlock2003} to the most general case where the coupling terms can be position-dependent. Inserting the modal expansion
\begin{equation}
\bigB (\rr) = \sum_\nu \left(
\left[
\begin{array}{c}
  u_{1\nu}(\rr) \\
  v_{1\nu}(\rr) \\
  u_{2\nu}(\rr) \\
  v_{2\nu}(\rr) \\
\end{array}
\right] \bh_\nu
+ \left[
\begin{array}{c}
  v^*_{1\nu}(\rr) \\
  u^*_{1\nu}(\rr) \\
  v^*_{2\nu}(\rr) \\
  u^*_{2\nu}(\rr) \\
\end{array}
\right] \bhd_\nu
\right)\,,
\label{eq:BogoliubovTransformation}
\end{equation}
with $\bh_\nu$ the annihilation operator of an elementary excitation of the coupled two-component Bose gas,
into Eq.~(\ref{eq:H2two}), we find
\begin{equation}
\hamGC^{(2)} = \frac{1}{2}\sum_\nu E_\nu \left( \bhd_\nu\bh_\nu + \bh_\nu\bhd_\nu \right)\,,
\label{eq:BogoDiag}
\end{equation}
provided that the wave functions fulfill the so-called coupled Bogoliubov equations:
\begin{equation}
\left[
\begin{array}{cc}
  \LGP_1 & \Gamma \\
\Gamma^* &\LGP_2\\
\end{array}
\right]
\left[
\begin{array}{c}
u_{1\nu}\\
v_{1\nu}\\
u_{2\nu}\\
v_{2\nu}
\end{array}
\right]
=E_\nu
\left[
\begin{array}{c}
u_{1\nu}\\
v_{1\nu}\\
u_{2\nu}\\
v_{2\nu}
\end{array}
\right]
\label{eq:EigenBogo}
\end{equation}
and the bi-orthogonality conditions
\begin{eqnarray}
&&
\sum_\sigma \int \dd\rr\ \Big[u_{\sigma\nu}(\rr)u^*_{\sigma\nu'}(\rr)-v_{\sigma\nu}(\rr)v^*_{\sigma\nu'}(\rr)\Big] = \delta_{\nu\nu'}
\label{eq:biorthogonal1} \\
&&
\sum_\sigma \int \dd\rr\ \Big[u_{\sigma\nu}(\rr)v_{\sigma\nu'}(\rr)-v_{\sigma\nu}(\rr)u_{\sigma\nu'}(\rr)\Big] = 0\,.
\label{eq:biorthogonal2}
\end{eqnarray}
These modes (indexed by $\nu$), being of bosonic nature, satisfy the Bose commutation rules
$[\bh_{\sigma\nu},\bhd_{\sigma'\nu'}]=\delta_{\sigma\sigma'}\delta_{\nu\nu'}$
and $[\bh_{\sigma\nu},\bh_{\sigma'\nu'}]=0$.

Notice that within this approach, we have disregarded the contribution of zero-mode terms in the modal
expansion~(\ref{eq:BogoliubovTransformation}).
The latter corresponds to two conjugate operators representing collective coordinates~\cite{mora2003}.
They induce quantum phase diffusion~\cite{lewenstein1996} and
fluctuations of the numbers of particles~\cite{mora2003}.
These effects are expected to be small in the limit of large numbers of particles that we consider here.

\subsubsection{Orthogonal field operator}
Another subtle issue of the present approach is that the normal terms $\hat{B}_\sigma (\rr)$ defined in
Eq.~(\ref{eq:BogoliubovTransformation}) do not fulfill the bosonic commutation relations.
As pointed out in Refs.~\cite{mora2003,castin1998}, the field operators $\hat{B}_\sigma (\rr)$ should
be orthogonalized with respect to the (quasi-)condensate wave function
$\psi_\sigma (\rr) \equiv \textrm{e}^{i\te_\sigma}\sqrt{\n_\sigma}$,
which amounts to apply the substitution $\hat{B}_\sigma (\rr) \rightarrow \hat{\Lambda}_\sigma (\rr)$ with
\begin{equation}
\hat{\Lambda}_\sigma (\rr)
\equiv
\hat{B}_\sigma (\rr) - \frac{{\psi}_\sigma (\rr)}{N_\sigma} \int \dd\rr^\prime\ \hat{B}_\sigma (\rr^\prime){\psi}_\sigma^* (\rr^\prime) \,.
\label{eq:orthogonalization}
\end{equation}
We then have
\begin{equation}
\hat{\Lambda}_\sigma (\rr) =
\sum_{\nu} \Big[ u_{\sigma\nu}^{\perp} (\rr) \hat{b}_{\nu} + v_{\sigma\nu}^{\perp*} (\rr) \hat{b}^\dagger_{\nu} \Big]
\label{eq:LambdaExp}
\end{equation}
with
\begin{eqnarray}
&& u_{\sigma\nu}^{\perp} \equiv u_{\sigma\nu} - \frac{{\psi}_\sigma (\rr)}{N_\sigma} \int \dd\rr^\prime\ u_{\sigma\nu} (\rr^\prime){\psi}_\sigma^* (\rr^\prime)
\label{eq:orthogonalization1} \\
&& v_{\sigma\nu}^{\perp} \equiv v_{\sigma\nu} - \frac{{\psi}_\sigma^* (\rr)}{N_\sigma} \int \dd\rr^\prime\ v_{\sigma\nu} (\rr^\prime){\psi}_\sigma (\rr^\prime) \,.
\label{eq:orthogonalization2}
\end{eqnarray}
According to Eqs.~(\ref{eq:commutBa}) and (\ref{eq:commutBb}), the orthogonal field operators $\hat{\Lambda}_\sigma$ satisfy the modified commutation rules  
\begin{eqnarray}
&& [{\hat{\Lambda}_{\sigma}}(\rr),{\hat{\Lambda}_{\sigma^\prime}}(\rr^\prime)] = 0
\label{eq:commutLa} \\
&& [{\hat{\Lambda}}_{\sigma}(\rr),{\hat{\Lambda}}^{\dagger}_{\sigma^\prime}(\rr^\prime)] = \delta_{\sigma\sigma^\prime} \left[ \delta(\rr \! - \! \rr^\prime) - \frac{{\psi}_\sigma (\rr){\psi}^*_\sigma (\rr^\prime)}{N_\sigma} \right].
\label{eq:commutLb}
\end{eqnarray}

The solutions of the non-Hermitian eigenvalue problem~(\ref{eq:EigenBogo}), together with the bi-orthogonality conditions~(\ref{eq:biorthogonal1}) and (\ref{eq:biorthogonal2}) and the orthogonalization process~(\ref{eq:orthogonalization1}) and (\ref{eq:orthogonalization2}), determine the excitation spectrum of the two-component Bose gas in the weakly-interacting regime. A mode $\nu$ describes a coupled two-component elementary excitation (Bogoliubov quasiparticle) of the mixture. The energy and wave functions of these excitations are $E_\nu$ and $\{u_{1\nu}^{\perp}(\rr),v_{1\nu}^{\perp}(\rr),u_{2\nu}^{\perp}(\rr),v_{2\nu}^{\perp}(\rr)\}$, respectively. They can be determined numerically, or, in certain cases, analytically. All physical observables can then be constructed by expansion on the corresponding basis.

\subsection{Correlation functions}
\label{sec:MFtheory.correlation}
We now consider the correlation properties of observable quantities, namely the phases and the densities of the two-component Bose gas. 
These quantities can be measured independently for each component in experiments with ultracold atoms,
using a gaseous mixture of a single bosonic atom prepared in two different internal states~\cite{myatt1997,matthews1998,hall1998a,hall1998b} and internal-state dependent imaging techniques~\cite{greiner2003}.
The density profiles, fluctuations and correlation functions of each component are then found directly from the images~\cite{esteve2006,armijo2010}.
The phase fluctuations and correlation functions of each component are found by time-of-flight~\cite{dettmer2001,hellweg2003} or Bragg spectroscopy~\cite{richard2003,gerbier2003,cacciapuoti2003} techniques.
The total and relative density profiles are then obtained by addition or subtraction of those of each component, which also provides their fluctuations and correlation functions.
Finally, the correlation function of the relative phase, $\te=\te_1-\te_2$, can be found using matter-wave interference techniques~\cite{hall1998b,betz2011}.

For each component $\sigma$, the phase correlation function is
\begin{eqnarray}
G_{\theta}^{\sigma}(\rr,\rr^\prime)
& \equiv & \la \hat{\te}_\sigma (\rr) \hat{\te}_\sigma (\rr^\prime) \ra -\la \hat{\te}_\sigma (\rr) \ra \la \hat{\te}_\sigma (\rr^\prime) \ra
\phantom{\frac{A}{A}} \nonumber \\
& = & -\frac{
\la:
( \Lh_{\sigma}-\Lh^{\dagger}_{\sigma} )
( \Lh^{\prime}_{\sigma}-\Lh^{\dagger\prime}_{\sigma} )
:\ra
}{4 \sqrt{n_\sigma \, n_\sigma^\prime}}\,,
\label{eq:theta-theta-s}
\end{eqnarray}
where the nude (resp.\ primed) quantities are evaluated at point $\rr$ (resp.\ $\rr^\prime$).
The operator $:~:$ represents normal ordering with respect to the orthogonal field operators $\Lh$ and $\Lh^\dagger$, which is used to avoid unphysical divergences~\cite{mora2003}.
Similarly, the density correlation function is
\begin{eqnarray}
G_{n}^{\sigma}(\rr,\rr^\prime)
& \equiv & \la \n_\sigma (\rr) \n_\sigma (\rr^\prime) \ra -\la \n_\sigma (\rr) \ra \la \n_\sigma (\rr^\prime) \ra
\phantom{\frac{A}{A}} \nonumber \\
& = & \sqrt{n_\sigma \, n^\prime_\sigma} \,
\la:
( \Lh_{\sigma}+\Lh^{\dagger}_{\sigma} )
( \Lh^{\prime}_{\sigma}+\Lh^{\dagger\prime}_{\sigma} )
:\ra\,.
\end{eqnarray}
Using the expansion of the orthogonal field operator into the basis of orthogonal Bogoliubov modes, Eq.~(\ref{eq:LambdaExp}), and the usual auxiliary wave functions~\footnote{Here, we use the notations
$f_{\sigma \nu}^{\textrm{p,m}}$ instead of the more usual notations
$f_{\sigma \nu}^{\pm}$ because the $\pm$ sign below labels a different quantity (the two branches of the spectra).}
\begin{eqnarray}
f_{\sigma \nu}^{\textrm{p}}(\rr) = u^\perp_{\sigma \nu}(\rr) - v^\perp_{\sigma \nu}(\rr)\,,
\label{eq:DefF1} \\
f_{\sigma \nu}^{\textrm{m}}(\rr) = u^\perp_{\sigma \nu}(\rr) + v^\perp_{\sigma \nu}(\rr)\,,
\label{eq:DefF2}
\end{eqnarray}
we then get the following explicit expressions after some algebraic calculations:
\begin{equation}
G_{\theta}^{\sigma}(\rr,\rr') =
\frac{1}{2 \sqrt{n_\sigma n_\sigma^\prime}}
\sum_\nu \mathcal{R}e
\Big[
  f_{\sigma \nu}^{\textrm{p}} f^{{\textrm{p}}\prime*}_{\sigma \nu} N_\nu
- f_{\sigma\nu}^{\textrm{p}} v^{\perp\prime*}_{\sigma \nu}
\Big]
\label{eq:PhasePhaseCorr}
\end{equation}
and
\begin{equation} 
G_{n}^{\sigma}(\rr,\rr') =
2 \sqrt{n_\sigma n_\sigma^\prime}
\sum_\nu \mathcal{R}e
\Big[
  f_{\sigma \nu}^{\textrm{m}} f^{{\textrm{m}}\prime*}_{\sigma \nu} N_\nu
+ f_{\sigma \nu}^{\textrm{m}} v^{\perp\prime*}_{\sigma \nu}
\Big],
\label{eq:DensDensCorr}
\end{equation}
where 
\begin{equation}
N_\nu = \frac{1}{\textrm{exp}({E_\nu/\kB T})-1}
\end{equation}
is the thermal population of mode $\nu$, according to the Bose-Einstein statistical distribution.
Note that expressions~(\ref{eq:PhasePhaseCorr}) and (\ref{eq:DensDensCorr}) are symmetric in $(\rr,\rr^\prime)$. This can be checked by noting that the commutation rule $[{\hat{\Lambda}_{\sigma}}(\rr),{\hat{\Lambda}_{\sigma}}(\rr^\prime)] = 0$ [see Eq.~(\ref{eq:commutLa})] implies the relation
$\sum_\nu u_{\sigma\nu}^\perp(\rr) v_{\sigma\nu}^{\perp *}(\rr^\prime) = \sum_\nu u_{\sigma\nu}^\perp(\rr^\prime) v_{\sigma\nu}^{\perp *}(\rr)$.

The two-point correlation function of the relative phase is defined by the same formula as Eq.~(\ref{eq:theta-theta-s}) with $\theta_\sigma$ replaced with $\theta=\theta_1-\theta_2$.
The same calculation strategy yields
\begin{eqnarray}
G_{\theta}(\rr,\rr^\prime) & \! = \! & \frac{1}{2}
\sum_{\nu}\mathcal{R}e\Big[\Big(\frac{f^{{\textrm{p}}}_{1 \nu}}{\sqrt{n_{1}}}-\frac{f^{{\textrm{p}}}_{2 \nu}}{\sqrt{n_{2}}}\Big)\Big(\frac{f^{{{\textrm{p}}}\prime}_{1\nu}}{\sqrt{n_{1}^\prime}}-\frac{f^{{{\textrm{p}}}\prime}_{2 \nu}}{\sqrt{n_{2}^\prime}}\Big)^* N_\nu
\nonumber \\ 
&& -\Big(\frac{f^{{\textrm{p}}}_{1 \nu}}{\sqrt{n_{1}}}-\frac{f^{{\textrm{p}}}_{2 \nu}}{\sqrt{n_{2}}}\Big)\Big(\frac{v^{\perp\prime}_{1 \nu}}{\sqrt{n_1^\prime}}-\frac{v^{\perp\prime}_{2 \nu}}{\sqrt{n_2^\prime}}\Big)^*\Big]\,.
\label{eq:relativePhaseCorr}
\end{eqnarray}

Having developed a general formalism for calculating the excitation modes of the two-component Bose gas with arbitrary one- and two-body couplings,
and established general formulas for the density and phase correlation functions, we explicitly calculate these quantities
in the homogeneous case in the next section.

\section{Homogeneous systems}
\label{sec:homog}
In this section, we consider a homogeneous system, where all potentials ($V_1$ and $V_2$) and coupling terms ($g_1$, $g_2$, $g_{12}$ and $\Omega$) in Hamiltonians~(\ref{eq:Ksigma}) and (\ref{eq:K12}) are independent of the position. Assuming that the potentials $V_1$ and $V_2$ are equal~\footnote{In the case where $V_1 \neq V_2$, the densities $n_1$ and $n_2$ would be modified compared to the following calculations. It is expected to lead to similar effects as those due to a modification of the coupling parameters $g_1$, $g_2$ and $g_{12}$.}, it can be assumed without loss of generality that $V_1=V_2=0$. This case allows for analytical calculations and contains the main physical effects discussed below.
Hereafter, we first rewrite the formalism of Sec.~\ref{sec:MFtheory} in a form adapted to the homogeneous case (Sec.~\ref{sec:homog.rewrite}). We then solve it in the most general situation where both one-body and two-body couplings coexist to discuss the excitation spectrum and wavefunctions, as well as density, phase, and relative-phase fluctuations of the two-component gas (Sec.~\ref{sec:homog.general}).

\subsection{Mean-field equations}
\label{sec:homog.rewrite}
Since all derivative terms in the Euler-Lagrange equations~(\ref{eq:GPE1}) and (\ref{eq:GPE2}) vanish in the homogeneous case, it immediately follows from Eq.~(\ref{eq:GPE2}) that $\theta-\alpha = 0$ or $\pi$ if $\Omega = \Omega_0 \textrm{e}^{-i\alpha} \neq 0$. Inserting these two solutions into the meanfield version of Eq.~(\ref{eq:K12BIS}), we find that $\theta = \alpha$ is a maximum of $E_\textrm{\tiny MF}$ and is thus an unstable solution. The stable solution is $\theta = \alpha + \pi$, which is a minimum of $E_\textrm{\tiny MF}$. For instance, the two components are in phase (resp.\ out of phase) when $\Omega \in \mathbb{R}^-$ (resp.\ $\Omega \in \mathbb{R}^+$).
If $\Omega = 0$, the relative phase $\theta$ is not a determined quantity as already discussed in the first paragraph of Sec.~\ref{sec:MFtheory.PhaseDens}.
Inserting the stable solution into Eq.~(\ref{eq:GPE1}), we then find
\begin{eqnarray}
g_1n_1+g_{12}n_2-\mu-\frac{\hbar\Omega_0}{2}\sqrt{\frac{n_2}{n_1}} &=& 0
\label{eq:GPE1b1} \\
g_2n_2+g_{12}n_1-\mu-\frac{\hbar\Omega_0}{2}\sqrt{\frac{n_1}{n_2}} &=& 0
\label{eq:GPE1b2}
\end{eqnarray}
and $n_1+n_2=n=N/\mathcal{V}$ with $N$ the total number of particles and $\mathcal{V}$ the volume of the system.
We assume that the parameters are such that the two components are miscible,
i.e.\ there exists a homogeneous solution of Eqs.~(\ref{eq:GPE1b1}) and (\ref{eq:GPE1b2}) of minimal energy with $n_1>0$ and $n_2>0$.

Translation invariance ensures that the Bogoliubov modes are the plane waves
\begin{eqnarray}
u_{\sigma \vk}(\rr) &=& \frac{1}{\sqrt{\mathcal{V}}}\tilde{u}_{\sigma \vk} e^{i\vk.\rr}
\label{eq:uk} \\
v_{\sigma \vk}(\rr) &=& \frac{1}{\sqrt{\mathcal{V}}}\tilde{v}_{\sigma \vk} e^{i\vk.\rr}\,,
\label{eq:vk} \\
f^\textrm{p/m}_{\sigma \vk}(\rr) &=& \frac{1}{\sqrt{\mathcal{V}}}\tilde{f}^{p/m}_{\sigma \vk} e^{i\vk.\rr}\,,
\label{eq:fk}
\end{eqnarray}
where we label the modes by the wave vector $\vk$ (instead of $\nu$).
In the following, we omit the tilde sign to simplify the notations.
Then, the amplitudes $u_{1 \vk}$, $v_{1 \vk}$, $u_{2 \vk}$, and $v_{2 \vk}$ are the solutions of the eigenproblem~(\ref{eq:EigenBogo})
for the diagonal blocks
\begin{equation}
\mathcal{L}^{GP}_\sigma =
\left[
\begin{array}{cc}
  +\textbf{A}_{\sigma\vk} & + g_\sigma n_\sigma \\
-g_\sigma n_\sigma & -\textbf{A}_{\sigma\vk}
\end{array}
\right], \
\label{eq:LGPsigma}
\end{equation}
with $\textbf{A}_{\sigma\vk} = \epsilon_{\vk}+2g_\sigma n_\sigma+g_{12}n_{\bar{\sigma}}-\mu$
where $\epsilon_{\vk} = \hbar^2\vk^2/2m$ is the free-particle dispersion relation,
and for the off-diagonal blocks
\begin{equation}
\Gamma=\left[
\begin{array}{cc}
 +g_{12}\sqrt{n_1n_2}-\hbar\Omega_0/2 & +g_{12}\sqrt{n_1n_2} \\
 -g_{12}\sqrt{n_1n_2} & -g_{12}\sqrt{n_1n_2}+\hbar\Omega_0/2
\end{array}
\right]\,.
\label{eq:Gammasigma}
\end{equation}
The biorthogonality conditions~(\ref{eq:biorthogonal1}) and (\ref{eq:biorthogonal2}) reduce to
\begin{equation}
\sum_{\sigma=1,2} \left( \vert u_{\sigma\vk}\vert^2 - \vert v_{\sigma\vk}\vert^2 \right) = 1\,
\label{eq:biorthogonalk}
\end{equation}
or equivalently
\begin{equation}
f^\textrm{m}_{1 \vk}f^\textrm{p}_{1 \vk}+f^\textrm{m}_{2 \vk}f^\textrm{p}_{2 \vk}=1,
\label{eq:biorthogonalkf}
\end{equation}
since the $f_{\sigma \vk}^{\textrm{p/m}}$ functions can be chosen to be real.
Note that since the classical fields $\phi_\sigma$ is homogeneous and the Bogoliubov wave functions $u_{\sigma\vk}$ and $v_{\sigma\vk}$ are plane waves, the orthogonalization procedure of Eqs.~(\ref{eq:biorthogonal1}) and (\ref{eq:biorthogonal2}) is irrelevant for $\vk \neq 0$.

Finally, the correlation functions introduced in Sec.~\ref{sec:MFtheory.correlation} are found by inserting Eqs.~(\ref{eq:uk}) and (\ref{eq:vk}) into Eqs.~(\ref{eq:PhasePhaseCorr}) and (\ref{eq:DensDensCorr}), which yields the following explicit formulas. For the phase correlation function of component $\sigma$,
\begin{equation} 
G_{\theta}^{\sigma}(\rr,\rr') = \frac{1}{2n_\sigma\mathcal{V}}
\sum_{\vk \neq 0} \Big[\vert f_{\sigma \vk}^{\textrm{p}} \vert^2 N_\vk - f_{\sigma \vk}^{\textrm{p}} v_{\sigma \vk}^*\Big]\! \cos \left[\vk.(\rr-\rr')\right].
\label{eq:PhaseCorrHomog}
\end{equation}
For the density correlation function of component $\sigma$,
\begin{equation} 
G_{n}^{\sigma}(\rr,\rr') = \frac{2n_\sigma}{\mathcal{V}}
\sum_{\vk \neq 0} \Big[\vert f_{\sigma \vk}^{\textrm{m}}\vert^2 N_\vk + f_{\sigma \vk}^{\textrm{m}} v_{\sigma \vk}^*\Big] \cos \left[\vk.(\rr-\rr')\right].
\label{eq:DensityCorrHomog}
\end{equation}
Similarly, the correlation function of the relative phase is
\begin{eqnarray}
G_{\theta}(\rr,\rr') && = \frac{1}{2\mathcal{V}}
\sum_{\vk \neq 0}
\Big[\Big\vert\frac{f_{ 1\vk}^{\textrm{p}}}{\sqrt{n_1}} \!-\! \frac{f_{2\vk}^{\textrm{p}}}{\sqrt{n_2}} \Big\vert^2 N_\vk 
\label{eq:RelativePhaseCorrHomog} \\
&& -\Big(\frac{f_{ 1\vk}^{\textrm{p}}}{\sqrt{n_1}} \!-\! \frac{f_{2\vk}^{\textrm{p}}}{\sqrt{n_2}} \Big)\Big(\frac{v_{1\vk}}{\sqrt{n_1}} \!-\! \frac{v_{2\vk}}{\sqrt{n_2}}\Big)^*\Big]\! \cos\left[\vk.(\rr-\rr')\right].
\nonumber
\end{eqnarray}
Notice that, for simplicity, we have indicated only $\vk \neq 0$ below the sum symbols of Eqs.~(\ref{eq:PhaseCorrHomog}), (\ref{eq:DensityCorrHomog}), and (\ref{eq:RelativePhaseCorrHomog}).
As a matter of fact, we will see that in general the Bogoliubov spectrum displays two branches,
over which the sums should be performed.

\subsection{Excitation spectrum and correlations}
\label{sec:homog.general}

We now study the excitation spectrum and the correlation functions of the homogeneous two-component Bose gas. 
Detailed calculations in the most general case are provided in Appendix~\ref{sec:A}. In brief, we generically find that the excitation spectrum is composed of two branches (see Fig.~\ref{fig:Case3spectrum}), one being gapped provided $\Omega_0\ne 0$, and the other one being ungapped and of Bogoliubov type. Both are particle-like at high energy. The two branches are found to be always distinct except if $\Omega_0=g_{12}=0$, in which case they both coincide with the usual Bogoliubov spectrum, $E_{\vk} = \sqrt{\epsilon_{\vk}\left(\epsilon_{\vk}+2\mu\right)}$. This holds for any positive values of $g_1$ and $g_2$.
For the sake of simplicity, we restrict in the following to the case where the two intra-component couplings are equal, $g_1=g_2$, which captures the main physics of the problem and is technically simpler. We assume that $g_{12}< g$, which is the miscibility condition for $\Omega_0=0$~\cite{ho1996}.

\subsubsection{Meanfield background and Bogoliubov excitations}
 
\begin{figure}[t!]
\includegraphics[width=8.7cm,clip=true]{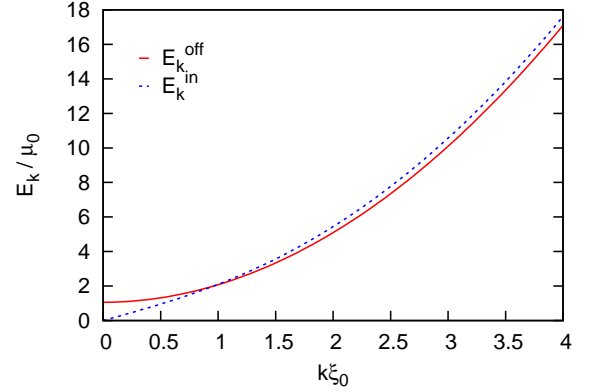}
\caption{\label{fig:Case3spectrum}
Bogoliubov spectrum of the coupled excitations in a homogeneous two-component Bose gas with $g_{12} \neq 0$ and $\Omega \neq 0$.
Plotted are the two energy branches $E_{\vk}^\textrm{in/off}$ [Eqs.~(\ref{eq:case3spectr1}) and~(\ref{eq:case3spectr2})] in the case $g_1=g_2$, for $g_{12}= 0.7 g_1 $ and $\hbar\Omega_0=0.4 g_1 n $. This corresponds to a situation where $g_{12}n>\hbar\Omega_0$ and the two branches cross at a certain momentum $k^{c}$ (see text). For $g_{12}n<\hbar\Omega_0$, there is no crossing point and the \offb is always above the \inb. Here, $\mu_0=g_1 N/2\mathcal{V}$ is the chemical potential in the absence of any coupling, and $\xi_0=\hbar/\sqrt{2m\mu_0}$ is the corresponding healing length.
}
\end{figure}

In the case $g_1=g_2\equiv g$, the meanfield densities of the two components are equal, $n_1=n_2$, and Eqs.~(\ref{eq:GPE1b1}) and~(\ref{eq:GPE1b2}) yield the chemical potential 
\begin{equation} 
\mu=(g+g_{12})n/2-\hbar\Omega_0/2,
\label{eq:muHomog}
\end{equation}
with $n=n_{1}+n_2$ the total density.
The excitation spectrum is computed in Appendix~\ref{sec:A2} [see Eq.~(\ref{eq:appenspecinoff}) together with Eqs.~(\ref{eq:appenAsym}) and~(\ref{eq:appenBsym})]. As mentioned above, it is composed of two branches, which explicitly read 
\begin{eqnarray}
\Ein &=& \sqrt{\epsilon_{\vk}\left(\epsilon_{\vk} + gn + g_{12}n \right)}
\label{eq:case3spectr1} \\
\Eof &=& \sqrt{\left(\epsilon_{\vk} + \hbar\Omega_0 \right)\left(\epsilon_{\vk} + \hbar\Omega_0 + (g - g_{12})n \right)}\,.
\label{eq:case3spectr2}
\end{eqnarray}
as a function of the problem parameters. The meaning of the labels $"\textrm{in}"$ and $"\textrm{off}"$ used to distinguish the two branches will become clear later. The spectrum is plotted in Fig.~\ref{fig:Case3spectrum}. 
The \inb shows the usual (ungapped) Bogoliubov-like dispersion relation : 
it is phonon-like for $\epsilon_\vk \ll g n, g_{12}n$ and 
$\Ein \simeq c \hbar k$
with $c=\sqrt{(g+g_{12}) n/2m}$ the sound velocity;
it is free-particle-like
for $\epsilon_\vk \gg g n, g_{12}n$ and
$\Ein \simeq \epsilon_\vk + (g+g_{12}) n /2$.
Conversely, the \offb is gapped, and free-particle-like in both low and high-energy limits, provided $\Omega_0\neq 0$:
for $\epsilon_\vk \ll (g-g_{12}) n, \hbar\Omega_0$, we have 
$\Eof \simeq  E_\textrm{gap} + \frac{2\hbar\Omega_0+(g- g_{12})n}{2\sqrt{\hbar\Omega_0(\hbar\Omega_0+(g- g_{12})n)}}\epsilon_\vk$ where $E_\textrm{gap}=\sqrt{\hbar\Omega_0 (\hbar\Omega_0+(g- g_{12})n)}$;
for $\epsilon_\vk \gg (g-g_{12}) n, \hbar\Omega_0$, we have
$\Eof \simeq \epsilon_\vk + \hbar\Omega_0 + (g- g_{12}) n /2$.
Thus, at low energy, the \offb is always above the "in" branch. At higher energy though, it depends on the strengths of the two couplings, since the two branches are separated by an energy $\Delta=\lim_{k\to \infty}(\Eof-\Ein)=\hbar\Omega_0-g_{12}n$. For attractive two-body coupling, $g_{12}<0$, we have $\Ein<\Eof$ for any momentum $\vk$, and the separation $\Eof-\Ein$ increases with both $\Omega_0$ and $g_{12}$. Therefore, attractive two-body coupling cooperates with one-body coupling. In contrast, repulsive two-body coupling, $g_{12}>0$, competes with one-body coupling and tends to decrease the separation between the branches. If the repulsive interactions are strong enough, $g_{12}n>\hbar\Omega_0$, the two curves exhibit a crossing point, above which $\Ein>\Eof$. This happens at the energy  $\epsilon_\vk^c\equiv(\hbar k^{c})^{2}/2m=\hbar\Omega_0 [\hbar\Omega_0+(g- g_{12})n]/2(g_{12}n-\hbar\Omega_0)$. When increasing the repulsive inter-component interactions, this crossing first appears at high momentum $k\approx \infty$, and then moves to lower momenta. 

In the particular case where $\Omega_0=0$, the \offb as well turns to be Bogoliubov-like; it is ungapped and phonon-like at low energy and $\Eof \simeq c \hbar k$
with $c=\sqrt{(g-g_{12}) n/2m}$ the sound velocity. In this case, which can be viewed as the limiting situation where the crossing of the two branches takes place at $k=0$, the \offb entirely lies above the \inb for $g_{12}<0$, and entirely below for $g_{12}>0$.

\begin{figure}[t!]
\includegraphics[width=8.7cm,clip=true]{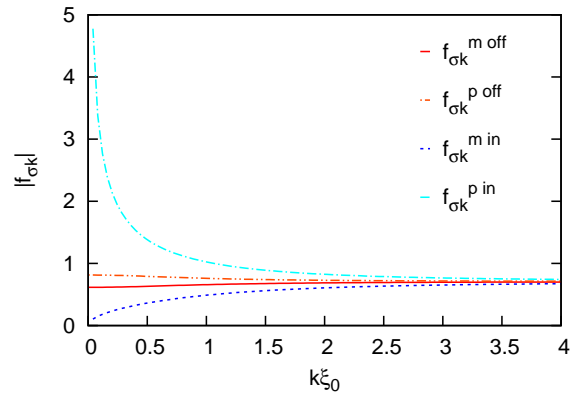}
\caption{\label{fig:Case3excitation}
Amplitudes of the wavefunctions $f_{\sigma\vk}^{\textrm{p/m}}$ of the coupled Bogoliubov excitations for a homogeneous two-component Bose gas with $g_{12} \neq 0$ and $\Omega \neq 0$.
Plotted are the absolute values, $\vert f_{\sigma\vk}^{\textrm{p/m}}\vert$ [see Eqs.~(\ref{eq:case3fmin}) to (\ref{eq:case3fpoff})] for the same parameters as in Fig.~\ref{fig:Case3spectrum}.
Since $g_1=g_2$, the absolute values are independent of the component $\sigma$.
The excitations are in phase in the \inb ($\Ein$) and off phase for the \offb ($\Eof$).
}
\end{figure}

Let us come back to arbitrary values of $\Omega_0$. The computation of the Bogoliubov wavefunctions is performed in the general case in the Appendix~\ref{sec:A1} [see Eqs.~(\ref{eq:appenf11}) to~(\ref{eq:appenf14})]. Their expressions in the case $g_1=g_2$ follow from the procedure indicated in Appendix.~\ref{sec:A2}, and read : 
\begin{eqnarray}
f_{1\vk}^{\textrm{m},\textrm{in}} = f_{2\vk}^{\textrm{m},\textrm{in}}&=&  \left[\frac{\epsilon_{\vk}}{2\Ein}\right]^{1/2}\label{eq:case3fmin} \\
f_{1\vk}^{\textrm{p},\textrm{in}} =f_{2\vk}^{\textrm{p},\textrm{in}} &=& \left[\frac{\Ein}{2\epsilon_{\vk}}\right]^{1/2} \label{eq:case3fpin} 
\end{eqnarray}
for the \inb, and
\begin{eqnarray}
f_{1\vk}^{\textrm{m},\textrm{off}}=-f_{2\vk}^{\textrm{m},\textrm{off}} &=&   \left[\frac{\epsilon_{\vk}+\hbar\Omega_0}{2\Eof}\right]^{1/2} \label{eq:case3fmoff} \\
f_{1\vk}^{\textrm{p},\textrm{off}}=-f_{2\vk}^{\textrm{p},\textrm{off}} &=&   \left[\frac{\Eof}{2\epsilon_{\vk}+2\hbar\Omega_0}\right]^{1/2}. \label{eq:case3fpoff}
\end{eqnarray}
for the "off" branch.
In the following, we omit the branch labels ("in"/"off") in the functions $f_{\sigma\vk}^{\textrm{p/m}}$ for simplicity, except when necessary. The moduli of the $f_{\sigma\vk}^{\textrm{p/m}}$ functions, which do not depend on the component $\sigma$ in the case $g_1=g_2$ considered here, are plotted in Fig.~\ref{fig:Case3excitation}.  
For the \inb, each component behaves as an effective single-component Bose gas with renormalized effective parameters, since the previous Bogoliubov spectrum and wavefunctions are similar to those of a single-component gas. Notice in particular the divergence of the $f_{\sigma \vk}^{\textrm{p}}$ functions. In contrast, the gapped dispersion relation of the \offb yields a different behavior for the $f_{\sigma \vk}^{\textrm{p/m}}$ functions. They do not depend much on $\vk$ as soon as $\hbar\Omega_0$ and $gn$ are of the same order, and in particular, the $f_{\sigma \vk}^{\textrm{p}}$ functions no longer diverge at low energy, since the gap acts as a low-momentum cut-off.\\

It follows as well from Eqs.~(\ref{eq:case3fmin}) to (\ref{eq:case3fpoff}) that, for a given component $\sigma$, the $f_\sigma^{\textrm{m}} (\rr)$ and $f_\sigma^{\textrm{p}} (\rr)$ wavefunctions are always in phase [i.e.\ $f_{\sigma\vk}^{\textrm{m}}f_{\sigma\vk}^{\textrm{p}}>0$]. 
Conversely, the modes associated to the components  $1$ ($f_{1\vk}^{\textrm{m}},f_{1\vk}^{\textrm{p}}$) and $2$ ($f_{2\vk}^{\textrm{m}},f_{2\vk}^{\textrm{p}}$) are off phase in the \offb and in phase in the "in" branch, hence the denomination used to label the two branches.
More precisely, since the separation $\Eof-\Ein$ increases with $\Omega_0$, we find that the one-body coupling $\Omega (\rr)$ tends to favor fluctuations of the phases of the components that are in phase, independently of its sign and more generally independently of its phase $\alpha$. This contrasts with the behavior of the mean-field phases $\theta_1$ and $\theta_2$, the difference of which is imposed by the phase of  $\Omega (\rr)$ (see Sec.~\ref{sec:homog.rewrite}). Indeed, the behavior of the fluctuations can by understood from the fact that the one-body coupling tends to impose the difference between the total phases of the two components. Since it is realized at the meanfield level, the phase fluctuations tend to be in phase, whatever the phase of $\Omega (\rr)$.
As regards two-body coupling, we find that $\Eof-\Ein$ decreases with $g_{12}$, so that for $g_{12}>0$, the two-body coupling favors off-phase density fluctuations whereas for $g_{12}<0$, it favors in-phase density fluctuations. This can be traced to the fact that for repulsive inter-component interactions ($g_{12}>0$), off-phase density fluctuations ($f_{1\vk}^{\textrm{m}} f_{2\vk}^{\textrm{m}} < 0$) cost less interaction energy than in-phase density fluctuations (and the other way round for $g_{12}<0$).
Therefore, for attractive two-body coupling, in-phase fluctuations are energetically favored, cooperatively by one-body and two-body couplings. Conversely, if the two-body coupling is repulsive and strong enough to compete with the one-body coupling ($g_{12}n>\hbar\Omega_0$), so that the two branches cross, they compete with the following result : for low-energy excitations ($\epsilon_\vk<\epsilon_\vk^c$), in-phase fluctuations cost less energy than off-phase fluctuations, whereas it is the opposite for high energy excitations ($\epsilon_\vk>\epsilon_\vk^c$).\\

\subsubsection{Fluctuations and correlations}

The phase and density correlations in each component $\sigma$ are determined by the $f_{\sigma \vk}^{\textrm{p}}$ and $f_{\sigma \vk}^{\textrm{m}}$ functions [see Eqs.~(\ref{eq:PhaseCorrHomog}) and (\ref{eq:DensityCorrHomog})].
Due to the similarity, in the in-phase branch, of the dispersion relation and formulas for the $f^{\textrm{p/m}}_{\sigma \vk}$ functions with those of a single-component Bose gas, each component behaves as an effective single-component gas.
The effective parameters, however, depend on all coupling parameters $g_1$, $g_2$ and $g_{12}$
and are, in general, different for the two components (if $g_1 \neq g_2$).
Then, the density fluctuations remain small for strong-enough interaction parameters and low temperatures in any dimension. In contrast, the behavior of the phase fluctuations strongly depends on the dimension, owing to the $1/\sqrt{\vert\vk\vert}$ divergence of the $f^\textrm{p,in}_{\sigma\vk}$ functions.
In three dimensions, the two components form true Bose-Einstein condensates with intra-component phase coherence. In lower dimensions, they form quasi-condensates with strong intra-component phase fluctuations driven by the ungapped Bogoliubov-like spectrum of the in-phase branch.

Let us turn to the relative phase correlations. Equation~(\ref{eq:RelativePhaseCorrHomog}) shows that in the case $g_1=g_2$ that we consider here, only the off-phase branch contributes to the sum. The correlation function for the relative phase can thus be rewritten
\begin{eqnarray}
G_{\theta}(\rr,\rr') = \frac{1}{2\mathcal{V}n}
\sum_{\vk \neq 0} && 
\Big\{\! 2 N_\vk + \Big(1 \! - \! \frac{\epsilon_\vk \! + \! \hbar\Omega_0}{\Eof}\Big) \!\Big\} \label{eq:Case3RelativePhaseCorrHomog}\\
&& \times \Big\vert f_{ 1\vk}^{\textrm{p},\textrm{off}}\!-\! f_{2\vk}^{\textrm{p},\textrm{off}} \Big\vert^2 \!\cos\left[\vk.(\rr-\rr')\right], \nonumber
\end{eqnarray}
making apparent the thermal and quantum contributions. Owing to the gap in the off-phase branch, its contribution remains finite, which ensures mutual phase coherence between the two Bose gases, in any dimension. This is, however, not true in the particular case $\Omega_0=0$, where the off-phase branch is ungapped: There, the two components are mutually phase coherent only in three dimensions, but show no true long-range mutual phase coherence in lower dimensions. Therefore, a finite one-body coupling suppresses the fluctuations of the relative phase,  
in agreement with the previous discussion according to which it tends to impose the phase at the meanfield level, favoring in-phase fluctuations of the phase.
To be more quantitative, we can rewrite Eq.~(\ref{eq:Case3RelativePhaseCorrHomog}) into the form
\begin{eqnarray}
G_\theta(\rr,\rr') = \frac{1}{n\mathcal{V}} \sum_{\vk \neq 0} & &
\Bigg[\sqrt{\frac{\epsilon_{\vk}+(g-g_{12})n+\hbar\Omega_0}{\epsilon_{\vk}+\hbar\Omega_0}} \times 
\label{eq:Case3RelativePhaseCoth}\\
&& \textrm{coth}\left(\!\frac{\Eof}{2\kB T}\!\right)-1\Bigg] \times \cos[\vk.(\rr - \rr')]\,.
\nonumber
\end{eqnarray}
Since $\Eof$ increases with $\Omega_0$ and both $\textrm{coth}(\Eof/2\kB T)$ and $\sqrt{(\epsilon_{\vk}+(g-g_{12})n+\hbar\Omega_0)/(\epsilon_{\vk}+\hbar\Omega_0)}$ decrease when $\Omega_0$ increases, the relative phase fluctuations $G_\theta(\rr,\rr)$ indeed decrease when the intensity of the one-body coupling increases. The influence of the two-body coupling on relative phase fluctuations is more involved. On the one hand, $\textrm{coth}\left(\!\frac{\Eof}{2\kB T}\!\right)$ is an increasing function of $g_{12}$ since $\Eof$ decreases when $g_{12}$ increases [see Eq.~(\ref{eq:case3spectr2})]. Indeed, an increase of the two-body coupling lowers the contributing off-phase branch, increasing its thermal occupancy. On the other hand, the amplitude of phase fluctuations in the off-phase branch, $\sqrt{\frac{\epsilon_{\vk}+(g-g_{12})n+\hbar\Omega_0}{\epsilon_{\vk}+\hbar\Omega_0}}=\Eof/(\epsilon_{\vk}+\hbar\Omega_0) \propto (f_{\vk}^{\textrm{p},\textrm{off}})^2$, is a decreasing function of $g_{12}$, which is intimately linked to the previously discussed observation that an increasing $g_{12}$ enhances the amplitude of off-phase density fluctuations.
To determine the overall behavior of the relative phase fluctuations, it is worth replacing $\sqrt{(\epsilon_{\vk}+(g-g_{12})n+\hbar\Omega_0)/(\epsilon_{\vk}+\hbar\Omega_0)}$ with $\Eof/(\epsilon_{\vk}+\hbar\Omega_0)$ in Eq.~(\ref{eq:Case3RelativePhaseCoth}).
Then, since $u\textrm{coth}(u)$ is an increasing function of $u$ (for $u>0$) and $\Eof$ is a decreasing function of $g_{12}$,
we conclude that the relative phase fluctuations decrease when the two-body coupling increases.
In particular, the relative phase fluctuations are maximally suppressed when $g_{12}>0$
approaches $g$ from below.
In other words, in a homogeneous two-component Bose gas, repulsive inter-component interactions reduce relative phase fluctuations while attractive inter-component interactions enhance relative phase fluctuations.\\

Let us mention that the physics of the general case $g_1 \neq g_2$ can be expected to be slightly different. Indeed, in this case, the contribution of the in-phase branch to the relative phase correlation function is non zero [see Eq.~(\ref{eq:RelativePhaseCorrHomog})] and the divergence of the $f_{\sigma \vk}^{\textrm{p},\textrm{in}}$ functions in this branch can lead to large fluctuations in low dimensions. Therefore, a small difference between $g_1$ and $g_2$ suppresses mutual phase coherence on large scales.\\

\begin{figure}[t!]
\includegraphics[width=7.cm,clip=true]{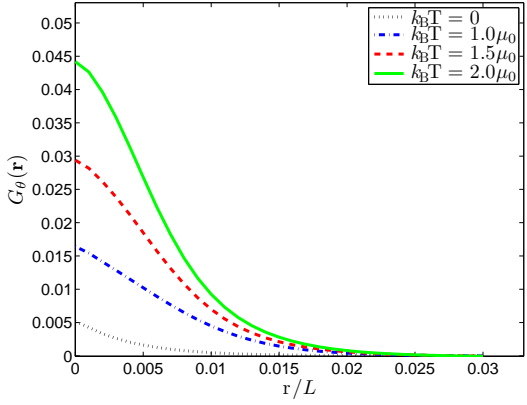}
\caption{\label{fig:Case3Correl}
Correlation function of the relative phase for a one-dimensional two-component Bose gas with one-body ($\Omega_0 \neq 0$) and two-body ($g_{12} \neq 0$) couplings, plotted for various temperatures ($\kB T/\mu_0=0$, $1$, $1.5$, $2$) in the case where $g_1=g_2\equiv g$. The parameters here  correspond to $N=10^4$ atoms of $^{87}$Rb ($m \simeq 144\times 10^{-27}$kg) in a 1D box of size $2L=10^{-4}$m, and interacting via the scattering length $a_1=a_2=5.95$nm. It corresponds in the absence of any coupling to the chemical potential $\mu_0=gn=7.88 \times 10^{-31}$J, which we choose as the energy unit. In these units, we use the parameters $\hbar\Omega_0=1 \mu_0$ and $g_{12}n=0.75 \mu_0$.
}
\end{figure}
Let us discuss as well the behavior of the relative phase correlation function $G_\theta(\rr,\rr')$ versus temperature, in the case $g_1=g_2$. Equation~(\ref{eq:Case3RelativePhaseCoth}) is plotted on Fig.~\ref{fig:Case3Correl} as a function of $|\rr-\rr'|$ for various temperatures, in the $1$D case. The function $G_\theta(\rr,\rr')$ generically decreases with $|\rr-\rr'|$ and goes to zero at large separations. Furthermore, it increases with the temperature $T$, as is easily checked from Eq.~(\ref{eq:Case3RelativePhaseCoth}), since the thermal contribution gets more and more important.

At zero temperature, the relative-phase correlation function reads
\begin{equation}
G_\theta(\rr) \! = \! \frac{1}{n} \int \frac{d\vk}{(2\pi)^d}
\Bigg[\frac{\Eof}{\epsilon_{\vk}+\hbar\Omega_0} -1 \Bigg]\cos(\vk.\rr),
\label{eq:Case3RelativePhaseZeroTInt}
\end{equation}
which is found by replacing the discrete sum in Eq.~(\ref{eq:Case3RelativePhaseCoth}) by an integral.
It can be seen from Eq.~(\ref{eq:case3spectr2}) that this function identically vanishes in the limit $g_{12}=g$. For $(g-g_{12})n\ll \hbar\Omega_0$, we can approximate $\frac{\Eof}{\epsilon_{\vk}+\hbar\Omega_0} -1$ by $\frac{(g-g_{12})n}{2(\epsilon_{\vk}+\hbar\Omega_0)}$ and analytically calculate the integral in Eq.~(\ref{eq:Case3RelativePhaseZeroTInt}).
In $1$D, it yields the exponentially decaying correlation function 
\begin{equation}
G_\theta^{1\textrm{D}}(\rr)=\frac{m(g-g_{12})n}{2n\hbar^{2}L_{\theta}^{-1}}e^{-|\rr|/L_{\theta}},
~~~\textrm{for}~T=0,
\label{eq:Case3RelativePhaseZeroTInt1D}
\end{equation}
where the correlation length is
\begin{equation}
L_\theta = \sqrt{\frac{\hbar}{2m\Omega_0}}.
\label{eq:Case3CorrLength}
\end{equation}
Equation~(\ref{eq:Case3RelativePhaseZeroTInt1D}) accurately reproduces the exact formula~(\ref{eq:Case3RelativePhaseZeroTInt}) plotted on Fig.~\ref{fig:Case3Correl},
which corresponds to $(g-g_{12})n = 0.25\hbar\Omega_0$.
In $3$D, we find
\begin{equation}
G_\theta^{3\textrm{D}}(\rr)=\frac{m(g-g_{12})n}{4\pi n\hbar^{2}|\rr|}e^{-|\rr|/L_{\theta}},
~~~\textrm{for}~T=0,
\label{eq:Case3RelativePhaseZeroTInt3D}
\end{equation}
which exhibits a divergence in $r=0$ and decreases over the same characteristic length $L_\theta$ as in 1D, Eq.~(\ref{eq:Case3CorrLength}).
For larger values of $(g-g_{12})n$, a formal expansion in powers of $(g-g_{12})n/\hbar\Omega_0$
of the term inside the brackets in Eq.~(\ref{eq:Case3RelativePhaseZeroTInt})
shows that the main dependence of the relative phase correlation function in $e^{-|\rr|/L_{\theta}}$ is preserved, with a multiplicative correction that is polynomial in $|\rr|/L_{\theta}$.
We numerically checked that the previous analytical formulas continue to hold up to this
polynomial correction in both $1$D and $3$D.
They predict in particular the correct correlation length, which therefore very weakly depends on the two-body coupling, although they tend to slightly overestimate the value of $G_\theta(0)$.

At finite temperature, the behavior of $G_\theta(\rr)$ at large separations $|\rr|$ can as well be obtained analytically. To do so, we replace in Eq.~(\ref{eq:Case3RelativePhaseCoth}) the discrete sum by an integral and use Eq.~(\ref{eq:case3spectr2}), which yields
\begin{equation}
G_\theta(\rr) \! = \! \frac{1}{n} \int \frac{d\vk}{(2\pi)^d}
\Bigg[\frac{\Eof}{\epsilon_{\vk}+\hbar\Omega_0} \textrm{coth}\left(\!\frac{\Eof}{2\kB T}\!\right)-1 \Bigg]\cos(\vk.\rr).
\label{eq:Case3RelativePhaseInt}
\end{equation}
The behavior at large $|\rr|$ is dominated by the components of momentum $k$ smaller than $1/r$. Thus, for $k_T|\rr|\gg 1$, where $k_T$ is defined by $E_{k_T}^{\textrm{off}}=\kB T$, we have $\kB T\gg \Eof$ for all contributing terms of the integral. Then, if $\kB T\gg (\epsilon_{\vk}+\hbar\Omega_0)$, Eq.~(\ref{eq:Case3RelativePhaseInt}) can be simplified to
\begin{eqnarray}
G_\theta(\rr) & \simeq &  \frac{1}{n} \int \frac{d\vk}{(2\pi)^d}
\frac{2\kB T}{\epsilon_{\vk}+\hbar\Omega_0} \cos(\vk.\rr).
\label{eq:Case3RelativePhaseInt2}
\end{eqnarray}
Notice that the previous condition requires that $\kB T\gg E_\textrm{gap},\, \hbar\Omega_0$, Eq.~(\ref{eq:Case3RelativePhaseInt2}) thus being valid in a large-separation and high-temperature regime. For $k_T|\rr|\gg 1$, the integral in Eq.~(\ref{eq:Case3RelativePhaseInt2}) can be calculated, yielding 
\begin{equation}
G_\theta^{1\textrm{D}}(\rr) \simeq \frac{2m\kB T}{n\hbar^{2}L_{\theta}^{-1}}e^{-|\rr|/L_{\theta}},
\label{eq:Case3CorrelApproximate1D}
\end{equation}
in one dimension, and
\begin{equation}
G_\theta^{3\textrm{D}}(\rr) \simeq \frac{m\kB T}{\pi n\hbar^{2}|\rr|}e^{-|\rr|/L_{\theta}},
\label{eq:Case3CorrelApproximate3D}
\end{equation}
in three dimensions.
Remarkably, we find the same expression for the correlation length of the relative phase [Eq.~(\ref{eq:Case3CorrLength})], as for zero temperature. In 1D, this result recovers that of Ref.~\cite{whitlock2003} and extends it to the case where one-body and two-body couplings coexist. The correlation length of the relative phase then weakly depends on the two-body coupling and decreases when the one-body coupling increases.
For smaller separations, the previous formulas no longer hold. A cut-off at $\vk_{T}$ in the integral would have to be taken into account, which in particular would solve the apparent divergence found in Eq.~(\ref{eq:Case3CorrelApproximate3D}) for $\rr=0$.

\begin{figure}[t!]
\includegraphics[width=8.5cm,clip=true]{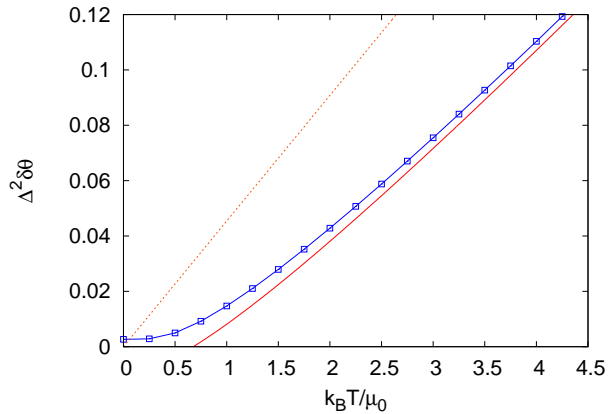}
\caption{\label{fig:Case3Fluct}
Relative phase fluctuations as a function of temperature for a one-dimensional two-component Bose gas with one-body ($\Omega_0 \neq 0$) and two-body ($g_{12} \neq 0$) couplings, plotted for the same parameters as in Fig.~\ref{fig:Case3Correl}.
The solid blue line with dots is the exact calculation, corresponding to Eq.~(\ref{eq:Case3RelativePhaseInt})
in $\rr=0$.
The solid red line is the expansion~(\ref{eq:FluctExpansion})
and the dotted red line corresponds to the first left-hand-side term in Eq.(\ref{eq:FluctExpansion}).
While the quantum fluctuations are small, the thermal contribution increases with temperature.
At high temperature, the exact calculation is accurately reproduced by
the high-temperature expansion~(\ref{eq:FluctExpansion}),
whereas the linear dominant term proves insufficient to do so.
}
\end{figure}

We finally discuss the temperature dependence of the relative phase fluctuations, which are given by $G_\theta(\rr=0)$. As already pointed out, the relative phase fluctuations always decrease with the one-body coupling $\Omega_0$, which thus favors mutual phase coherence between the two condensates. Moreover, repulsive two-body coupling ($g_{12}>0$) tends to reduce the fluctuations of the relative phase while attractive two-body coupling enhances them. The temperature dependence of those fluctuations is shown in Fig.~\ref{fig:Case3Fluct}, for the 1D case.
The zero-temperature fluctuations, which are given by their quantum contribution, are smaller than those of a single condensate~\cite{whitlock2003}.
The fluctuations then unsurprisingly increase with temperature. 
Their high-temperature behavior can be obtained by an analytical expansion, which we detail in the Appendix~\ref{sec:B}.
We find that for $\kB T\gg \hbar\Omega_0, (g-g_{12})n$, the dominant term is linear in $T$ and reads 
${2m\kB T}/{n\hbar^{2}L_{\theta}^{-1}}$, 
which coincides with the prefactor in Eq.~(\ref{eq:Case3CorrelApproximate1D}), and the result of~\cite{whitlock2003}. In particular, the one-body coupling favors local mutual phase coherence between the two components.
However, the dominant contribution is generally not sufficient to accurately reproduce the exact calculations
as shown in Fig.~\ref{fig:Case3Fluct}. In order to get a better accuracy, we include the next-order
contribution, which scales as $\sqrt{T}$ and, remarkably, is independent of the couplings.
More precisely, we find the high-temperature expansion
\begin{equation}
G_\theta(\rr=0)\simeq \frac{2m\kB T}{n\hbar^{2}L_{\theta}^{-1}}-\frac{I_1}{\pi}\sqrt{\frac{2\kB T}{\hbar^{2}n^2/2m}} + \textrm{O}(1),
\label{eq:FluctExpansion}
\end{equation}
where $I_1=\int_0^\infty du\ [1/u^2-\coth(u^2)-1 ] \simeq 1.82$.
As can be seen in Fig.~\ref{fig:Case3Fluct}, Eq.~(\ref{eq:FluctExpansion}) provides a fair approximation
to the exact calculations. In particular, we find that the $\sqrt{T}$ correction significantly
lowers the relative phase fluctuations.

\section{Conclusions}
\label{sec:conclusion}
In this paper, we have derived a general meanfield theory for a two-component Bose gas in the presence of both one-body and two-body couplings. We considered the most general situation where both one-body and two-body couplings can be position dependent, and where the gas can experience a component-dependent external potential. Our formulation uses the phase-density formalism, which allows us to capture both cases of true condensates and quasi-condensates with large phase fluctuations. We have written the coupled Gross-Pitaevskii equations, which determine the ground-state background, as well as the Bogoliubov equations, which determine the pair-excitation spectrum of the mixture. We obtained general formulas for phase and density correlation functions within each component, as well as for their relative phase, at zero and finite temperature.

We have then applied our formalism to a homogeneous case where both one-body and two-body couplings coexist (Sec.~\ref{sec:homog.general}). Our discussion then focused on the excitation spectrum and the relative phase fluctuations in the case of equal intra-component interactions, which captures the main physics. We summarize our main results in the following.

The excitation spectrum is composed of two branches, which are distinct provided at least one of the couplings is present. The first branch, which corresponds to in-phase fluctuations of the two Bose gases, is of Bogoliubov type. It depends only on the two-body coupling while being unaffected by one-body coupling. The second branch, which corresponds to off-phase fluctuations, is gapped as soon as the one-body coupling is non zero. The two branches cross each other at a given momentum if the two-body coupling is repulsive and exceeds the one-body coupling.

As regards phase and density fluctuations, each component behaves as an effective single-component Bose gas with coupling parameters that are renormalized by the inter-species two-body coupling.
In particular, while the density fluctuations remain small in all dimensions, the two components exhibit strong intra-component phase fluctuations in low dimensions, driven by the ungapped Bogoliubov-like spectrum of the in-phase branch.

The behavior of the relative phase is more involved. At the meanfield level, it is imposed by the one-body coupling, and in particular by its phase. Then, the fluctuations of the relative phase depend only on the modulus of the one-body coupling and on the two-body coupling.
At variance with the phase and density fluctuations within each component, the relative-phase fluctuations are mostly determined by the off-phase branch of the spectrum, provided that the intra-species interaction strengths are not too different. This is strictly the case where they are equal ($g_1=g_2$). Then, the two component are mutually phase coherent in any dimension, due to the gap in the contributing off-phase branch (provided $\hbar\Omega_0 \neq 0$).   
Therefore, the one-body coupling always favors relative-phase coherence of the two Bose gases, independently of its phase. 
As regards the two-body coupling, two mechanisms compete. On the one hand, an increasing $g_{12}$ tends to lower the contributing off-phase branch, hence increasing its thermal occupancy. On the other hand, it enhances the amplitude of off-phase density fluctuations, and therefore reduces the amplitude of phase fluctuations in the contributing off-phase branch. We found that the latter effect always dominates. Therefore, repulsive inter-component interactions suppress relative phase fluctuations while attractive inter-component interactions enhance relative phase fluctuations. 
Then, repulsive two-body coupling cooperates with one-body coupling and further suppresses relative-phase fluctuations, while attractive two-body coupling competes with one-body coupling and enhances relative-phase fluctuations. 
Closed analytical forms were eventually found for the relative-phase correlation function, in the high-temperature and large-separation regime. This enabled us to identify a correlation length for the relative phase, which was found to decrease when the one-body coupling increases, and to be roughly independent of the two-body coupling.

Our work generalizes previous results to the case where both one-body and two-body couplings are present between the two Bose components. The homogeneous cases we have analyzed are expected to contain the main physics of relative-phase coherence.
The formalism that we have developed here can be directly applied to more complicated situations. For instance, the effect of inhomogeneous trapping, which can be component-dependent, is particularly relevant in the context of ultracold-atom systems. In this case, one may resort to numerical solutions of the Gross-Pitaevskii and Bogoliubov equations.
Other interesting applications of this formalism include the study of the effects of strong inhomogeneities in interacting Bose gases, in particular random couplings, which is attracting much attention in ultracold-atom systems~\cite{lsp2010}. One may envision several applications.
First, disordered potentials have been shown to induce Anderson localization of the Bogoliubov excitations in single-component Bose gases~\cite{lugan2007b,gaul2008,gaul2011b,lugan2011}. How does it extend to the case of coupled Bose gases~?
Second, disorder can be included in interaction terms using inhomogeneous Feshbach resonances~\cite{gimperlein2005}. What would be the effect of random inter-species coupling~?
Third, disorder can be included in one-body coupling, which has been shown to produce random-field-induced-order of the relative phase of two Bose-Einstein condensates at zero temperature~\cite{wehr2006,niederberger2008,niederberger2009,niederberger2010}. How does finite temperature affect this behavior~?

\textit{Note added.}~Recently, we were made aware of a related work, reporting the analysis of the excitation spectrum and the structure factors of coupled two-component Bose-Einstein condensates~\cite{abad2013}.

\acknowledgments
This research was supported by
the European Research Council (FP7/2007-2013 Grant Agreement No.\ 256294),
the Agence Nationale de la Recherche (Contract No.\ ANR-08-blan-0016-01),
RTRA-Triangle de la Physique, and the Institut Francilien de Recherche sur les Atomes Froids (IFRAF).
We acknowledge the use of the computing facility cluster GMPCS of the 
LUMAT federation (FR LUMAT 2764) and assistance of M.~Besbes.


\bigskip

\begin{appendix}
\section{General formulas for the homogeneous two-component Bose gas}
\label{sec:A}
In this appendix, we compute the excitation spectrum and wavefunctions of the homogeneous two-component Bose gas in the most general situation where both one-body and two-body couplings are present. 

\subsection{General case, $g_1 \neq g_2$}
\label{sec:A1}
In principle, the first step is to solve the meanfield background, Eqs.~(\ref{eq:GPE1b1}) and (\ref{eq:GPE1b2}). However, in the most general case with $g_1 \neq g_2$, $\Omega_0 \neq 0$, and $g_{12} \neq 0$, we did not find a simple closed solution~\footnote{In the case $g_1=g_2\equiv g$, we have $n_1=n_2=N/2\mathcal{V}$, with $N$ the total number of atoms, by symmetry of the two components. Equations~(\ref{eq:GPE1b1}) and (\ref{eq:GPE1b2}) are then identical and yield the simple solution $\mu=(g+g_{12})N/2\mathcal{V}-\hbar\Omega_0/2$.},~\footnote{In the case $\Omega_0=0$, Equations~(\ref{eq:GPE1b1}) and (\ref{eq:GPE1b2}) reduce to a linear problem whose solution reads $n_\sigma=\frac{N}{\mathcal{V}}\frac{g_{\bar{\sigma}}-g_{12}}{g_1+g_2-2g_{12}}$ and $\mu=\frac{N}{\mathcal{V}}\frac{g_1g_2-g^2_{12}}{g_1+g_2-2g_{12}}$ with $\bar{\sigma}=2$ (resp.\ $1$) for $\sigma = 1$ (resp.\ $2$)}. Thus, in the following, we write directly the Bogoliubov equations as a function of $n_1$, $n_2$ and $\mu$.

Given the meanfield solution $n_1$, $n_2$ and $\mu$, one has to solve the homogeneous Bogoliubov equations~(\ref{eq:EigenBogo}) together with~(\ref{eq:LGPsigma}) and~(\ref{eq:Gammasigma}). 
By taking the sum and difference of the first two rows on the one hand, and of the last two rows of the other hand, we can rewrite those Bogoliubov equations in terms of the  $f_{\sigma\vk}^{\textrm{p,m}}$ functions :
\begin{eqnarray}
E_\vk f_{\sigma\vk}^{\textrm{m}}& = & \left( \epsilon_{\vk}+\dfrac{\hbar\Omega_0}{2}\sqrt{\frac{n_{\bar{\sigma}}}{n_{\sigma}}}\right) f_{\sigma\vk}^{\textrm{p}}-\dfrac{\hbar\Omega_0}{2} f_{\bar{\sigma}\vk}^{\textrm{p}} 
\label{eq:appenf1} \\
E_\vk f_{\sigma\vk}^{\textrm{p}}& = & \left( \epsilon_{\vk}+\dfrac{\hbar\Omega_0}{2}\sqrt{\frac{n_{\bar{\sigma}}}{n_{\sigma}}}+2g_\sigma n_\sigma\right) f_{\sigma\vk}^{\textrm{m}} \nonumber\\
 & & +\left( 2g_{12}\sqrt{n_1n_2}-\dfrac{\hbar\Omega_0}{2}\right) f_{\bar{\sigma}\vk}^{\textrm{m}}, 
\label{eq:appenf2} 
\end{eqnarray}
where $\bar{\sigma}$ is the conjugate of component $\sigma$ [$\bar{\sigma}=2$ (resp.\ $1$) for $\sigma = 1$ (resp.\ $2$)]. Using the normalization condition~(\ref{eq:biorthogonalkf}), it yields 
\begin{eqnarray}
E_\vk^2 f_{\sigma\vk}^{\textrm{p}}&=&\left( \epsilon_{\sigma\vk}+2U_\sigma \right) 
\left( \epsilon_{\sigma\vk}f_{\sigma\vk}^{\textrm{p}}-\dfrac{\hbar\Omega_0}{2}
 f_{\bar{\sigma}\vk}^{\textrm{p}}\right) 
 \label{eq:appenf3} \\
 & & +\left( 2U_{12}-\dfrac{\hbar\Omega_0}{2}\right) \left(-\dfrac{\hbar\Omega_0}{2}
 f_{\sigma\vk}^{\textrm{p}}+ \epsilon_{\bar{\sigma}\vk} f_{\bar{\sigma}\vk}^{\textrm{p}}\right)  \nonumber\\
E_\vk & =  &  f_{1\vk}^{\textrm{p}}\left(\epsilon_{1\vk}f_{1\vk}^{\textrm{p}}-\dfrac{\hbar\Omega_0}{2}f_{2\vk}^{\textrm{p}}\right) \nonumber \\
& & +f_{2\vk}^{\textrm{p}}\left(\epsilon_{2\vk}f_{2\vk}^{\textrm{p}}-\dfrac{\hbar\Omega_0}{2}f_{1\vk}^{\textrm{p}}\right),
\label{eq:appenf4}
\end{eqnarray}
where we have defined $\epsilon_{\sigma\vk}\equiv\epsilon_{\vk}+\dfrac{\hbar\Omega_0}{2}
\sqrt{\frac{n_{\bar{\sigma}}}{n_{\sigma}}}$, $U_\sigma\equiv g_\sigma n_\sigma$, and $U_{12}\equiv g_{12}\sqrt{n_1n_2}$. 
By defining as well
\begin{eqnarray}
A_{\vk\sigma}=\epsilon_{\sigma \vk}(\epsilon_{\sigma \vk}+2U_\sigma )-\dfrac{\hbar\Omega_0}{2}\left( 2U_{12}-\dfrac{\hbar\Omega_0}{2}\right)
\label{eq:appenAk}\\
B_{\vk\sigma}=\epsilon_{\bar{\sigma}\vk}\left( 2U_{12}-\dfrac{\hbar\Omega_0}{2}\right) -\dfrac{\hbar\Omega_0}{2}(\epsilon_{{\sigma}\vk}+2U_{{\sigma}})
\label{eq:appenBk}
\end{eqnarray}
we can rewrite Eq.~(\ref{eq:appenf3}) separating the terms in $f_{\sigma\vk}^{\textrm{p}}$ from those in $f_{\bar{\sigma}\vk}^{\textrm{p}}$
\begin{equation}
f_{\bar{\sigma}\vk}^{\textrm{p}} B_{\vk\sigma}= f_{\sigma\vk}^{\textrm{p}}[E_\vk^2-A_{\vk\sigma}]
\label{eq:appenf5} 
\end{equation}
The Bogoliubov energies are then found from the ratio of the two avatars of Eq.~(\ref{eq:appenf5}) corresponding to $\sigma=1$ and $\sigma=2$, respectively. It yields 
\begin{equation}
E_\vk^{\pm}=\sqrt{\frac{1}{2}(A_{\vk1}+A_{\vk2}) \pm \sqrt{(A_{\vk1}-A_{\vk2})^2/4+B_{\vk1}B_{\vk2}}}.
\label{eq:appenspec}
\end{equation}
The excitation spectrum is composed of two branches, the one labeled by $(+)$ always being above the one labeled by $(-)$. Their low- and high-momentum behaviors are easily found from a low- and high-momentum expansion of the $A_{k\sigma}$ and $B_{k\sigma}$. At low momentum, the $(-)$ branch is ungapped and phonon-like; conversely, the $(+)$ branch exhibits a finite gap as soon as $\Omega_0 \neq 0$, given by 
\begin{eqnarray}
E_\textrm{gap} & = &
\bigg[\frac{\hbar^2\Omega_0^2}{4}\left(2+\frac{n_1}{n_2} +\frac{n_2}{n_1}\right)
\label{eq:appengap} \\
& &  +\hbar\Omega_0\sqrt{n_1 n_2}(g_1+g_2-2g_{12})\bigg]^{1/2}.
\nonumber
\end{eqnarray}
At high energy, both branches are particle-like, and separated by an energy 
\begin{eqnarray}
\Delta & = &
\bigg[
\left( \dfrac{\hbar\Omega_0}{2}\dfrac{n_2-n_1}{\sqrt{n_1 n_2}}+g_1 n_1-g_2 n_2 \right)^2
\label{eq:appensep} \\
&& +(2g_{12}\sqrt{n_1 n_2}-\hbar\Omega_0)^2 \bigg]^{1/2}.
\nonumber
\end{eqnarray}
In between, the two branches can possibly coincide at a specific $\vk$ provided the equation $(A_{\vk1}-A_{\vk2})^2/4+B_{\vk1}B_{\vk2}=0$ has a solution (see Sec.~\ref{sec:homog.general} for a precise example in the case $g_1=g_2$).

In the particular case where $\Omega_0=g_{12}=0$, and only in this case~\footnote{For the two branches to be identical, one necessarily have $\Omega_0=0$ to make the $(+)$ branch ungapped [see Eq.~(\ref{eq:appengap})], and then $g_{12}=0$ to make the two branches coincide at high energy [see Eq.~(\ref{eq:appensep})].}, the two branches are identical and correspond to the usual single-particle Bogoliubov spectrum, $E_{\vk}^{{\pm}} = \sqrt{\epsilon_{\vk}\left(\epsilon_{\vk}+2\mu\right)}$. Notice that this holds even for $g_1 \neq g_2$ because the meanfield background is identical for the two Bose gases,
i.e.\ $g_1 n_1 = g_2 n_2 = \mu$ [see Eqs.~(\ref{eq:GPE1b1}) and (\ref{eq:GPE1b2}) with $\Omega_0=g_{12}=0$]. 
In this case, the spectrum shows twofold degeneracy (there is also a trivial $+\vk \leftrightarrow -\vk$ degeneracy, which we disregard here). 

Given the excitation spectrum, we can then compute the Bogoliubov wavefunctions $f_{\sigma\vk}^{\textrm{p,m}}$. To do so, we use Eq.~(\ref{eq:appenf5}) and express $f_{2\vk}^{\textrm{p}}$ as a function of $f_{1\vk}^{\textrm{p}}$. Inserting this expression into Eq.~(\ref{eq:appenf4}), we find 
\begin{equation}
f_{1\vk}^{\textrm{p}}=\sqrt{\dfrac{E_\vk}{\epsilon_{1\vk}-\hbar\Omega_0\dfrac{E_\vk^2-A_{\vk 1}}{B_{\vk 1}}+\epsilon_{1\vk}\left(\dfrac{E_\vk^2-A_{\vk 1}}{B_{\vk 1}}\right)^2}}
\label{eq:appenf11}
\end{equation}
up to an arbitrary phase that we set to zero. Using again Eq.~(\ref{eq:appenf5}), we find :
\begin{eqnarray}
f_{2\vk}^{\textrm{p}} & = & \dfrac{E_\vk^2-A_{\vk 1}}{B_{\vk 1}} \times
\label{eq:appenf12} \\
&& \sqrt{\dfrac{E_\vk}{\epsilon_{1\vk}-\hbar\Omega_0\dfrac{E_\vk^2-A_{\vk 1}}{B_{\vk 1}}+\epsilon_{1\vk}\left(\dfrac{E_\vk^2-A_{\vk 1}}{B_{\vk 1}}\right)^2}}.
\nonumber
\end{eqnarray}
Notice that although $f_{2\vk}^{\textrm{p}}$ could also be expressed by a symmetric expression as Eq.~(\ref{eq:appenf11}), this would not be sufficient to determine its relative phase with respect to  $f_{1\vk}^{\textrm{p}}$. We finally deduce the $f_{\sigma\vk}^{\textrm{m}}$ waves from the $f_{\sigma\vk}^{\textrm{p}}$ using Eq.~(\ref{eq:appenf1}). It yields
\begin{eqnarray}
f_{1\vk}^{\textrm{m}}=\dfrac{\epsilon_{1\vk}-\hbar\Omega_0(E_\vk^2-A_{\vk 1})/2B_{\vk 1}}{\sqrt{E_\vk \left[\epsilon_{1\vk}-\hbar\Omega_0\dfrac{E_\vk^2-A_{\vk 1}}{B_{\vk 1}}+\epsilon_{1\vk}\left(\dfrac{E_\vk^2-A_{\vk 1}}{B_{\vk 1}}\right)^2 \right]}} \nonumber \\
\label{eq:appenf13} 
\end{eqnarray}
and
\begin{eqnarray}
f_{2\vk}^{\textrm{m}}=\dfrac{\epsilon_{2\vk}(E_\vk^2-A_{\vk 1})/B_{\vk 1}-\hbar\Omega_0/2}{\sqrt{E_\vk \left[\epsilon_{1\vk}-\hbar\Omega_0\dfrac{E_\vk^2-A_{\vk 1}}{B_{\vk 1}}+\epsilon_{1\vk}\left(\dfrac{E_\vk^2-A_{\vk 1}}{B_{\vk 1}}\right)^2 \right]}}. \nonumber \\
\label{eq:appenf14} 
\end{eqnarray}

\subsection{Symmetric case, $g_1=g_2$}
\label{sec:A2}

In the case discussed in Sec.~\ref{sec:homog.general} where the intra-component couplings are equal, $g_1=g_2$, we have by symmetry of the two components $n_1=n_2$, $A_{\vk1}=A_{\vk2}\equiv A_{\vk}$, and $B_{\vk1}=B_{\vk2}\equiv B_{\vk}$, with
\begin{eqnarray}
A_{\vk} & = & \left(\epsilon_{\vk}+\dfrac{\hbar\Omega_0}{2}\right)\left(\epsilon_{\vk}+\dfrac{\hbar\Omega_0}{2}+gn\right) \nonumber \\
& & -\dfrac{\hbar\Omega_0}{2}\left( ng_{12}-\dfrac{\hbar\Omega_0}{2}\right) 
\label{eq:appenAsym} \\
B_{\vk}& = & \left(\epsilon_{\vk}+\dfrac{\hbar\Omega_0}{2}\right)\left( ng_{12}-\dfrac{\hbar\Omega_0}{2}\right)\nonumber \\ 
& & -\dfrac{\hbar\Omega_0}{2}\left(\epsilon_{\vk}+\dfrac{\hbar\Omega_0}{2}+gn\right).
\label{eq:appenBsym} 
\end{eqnarray}
Equation~(\ref{eq:appenspec}) then reads $E_\vk^{\pm}=\sqrt{A_{\vk} \pm |B_{\vk}|}$. Therefore, the two energies corresponding to a given momentum $\vk$, irrespective to the branches, are nothing but $\sqrt{A_{\vk} \pm B_{\vk}}$. This allows for redefining the two branches of the spectrum in a different way : 
\begin{equation}
E_\vk^{\textrm{in}/\textrm{off}}=\sqrt{A_{\vk} \pm B_{\vk}}
\label{eq:appenspecinoff}
\end{equation}
Although none of the branches is now systematically above or below the other one, this convention for the \inb and the \offb will prove more convenient in Sec.~\ref{sec:homog.general}, especially while computing the Bogoliubov wavefunctions. Indeed, notice that $(E_\vk^2-A_{\vk 1})/B_{\vk 1}=1$ for the \inb and $-1$ for the \offb. This enables us to considerably simplify Eqs.~(\ref{eq:appenf11}) to~(\ref{eq:appenf14}) for the Bogoliubov wavefunctions in the case $g_1 = g_2$.

\section{High-temperature expansion for the one-dimensional fluctuations of the relative phase}
\label{sec:B}
We perform here a high-temperature expansion of the relative phase fluctuations in the 1D case,
valid for $\kB T\gg \hbar\Omega_0, (g-g_{12})n$.
The relative-phase fluctuations are given by  
\begin{eqnarray}
G_\theta(0) & = &  \frac{1}{n\pi} \int_0^\infty dk
\Bigg[\frac{\Eof}{\epsilon_{\vk}+\hbar\Omega_0} \textrm{coth}\left(\!\frac{\Eof}{2\kB T}\!\right)-1 \Bigg], \nonumber\\
\label{eq:appenFluct}
\end{eqnarray}
with $\Eof=\sqrt{\left(\epsilon_{\vk} + \hbar\Omega_0 \right)\left(\epsilon_{\vk} + \hbar\Omega_0 + (g - g_{12})n \right)}$, see Eqs.~(\ref{eq:case3spectr2}) and (\ref{eq:Case3RelativePhaseInt}).

\subsection{General expansion and leading term}
Introducing $k_T$ such that $E_{k_T}^{\textrm{off}}=\kB T$, we can split the integral in 
Eq.~(\ref{eq:appenFluct}) into two parts, corresponding to $k<k_T$ and to $k>k_T$, respectively.
For $k\gg k_T$, $\textrm{coth}\left(\!\frac{\Eof}{2\kB T}\!\right)\approx 1$
up to some exponentially decaying terms.
Hence, we can safely approximate the first part of the integral by
$\frac{1}{n\pi} \int_{k_T}^\infty dk \left(\frac{\Eof}{\epsilon_{\vk}+\hbar\Omega_0} -1 \right)$,
the leading-order term of which scales as $1/k_T \propto 1/\sqrt{T}$ in the high-temperature limit.
We can thus disregard this contribution.
For $k\ll k_T$, we have $E_{k_T}^{\textrm{off}}\ll 2\kB T$ 
so that we can use the expansion 
$\coth(x)\approx_{x\rightarrow 0} 1/x+x/3-x^3/45+...$, yielding the contribution
\begin{equation}
\frac{1}{n\pi} \int_0^{k_T} dk
\Bigg[\frac{2\kB T}{\epsilon_{\vk}+\hbar\Omega_0} + \frac{\epsilon_{\vk} + \hbar\Omega_0 + (g - g_{12})n}{6\kB T}-...-1 \Bigg],
\label{eq:appenInt0Kt}
\end{equation}
where we have retained the first two contributions.
At high temperature, the first term is linear in $T$ and reads
$\frac{2m\kB T}{n\hbar^{2}L_{\theta}^{-1}}$, where $L_\theta=\sqrt{\frac{\hbar}{2m\Omega_0}}$. 
Then, all terms coming from the expansion of the $\coth$ function are of order $\sqrt{T}$ and more, and the last term coming from the $-1$ is constant. Therefore, at high temperature, the relative phase fluctuations scale linearly with $T$ :
\begin{equation}
G_\theta(\rr=0)\simeq \frac{2m\kB T}{n\hbar^{2}L_{\theta}^{-1}} + O(\sqrt{T})
\label{eq:appenExpansionPartielle}
\end{equation}
Obtaining the next correcting terms, scaling as $\sqrt{T}$, from Eq.~(\ref{eq:appenInt0Kt}) is not straightforward
since one would have to evaluate all terms of the integral and resum them.
Furthermore, with this approach, each term would depend on $k_T$,
which was introduced as a typical bound to split the integral and is thus somehow defined up to an
arbitrary constant of the order of one. It would prevent us to extract the correct numerical prefactor of the $\sqrt{T}$ term.

\subsection{Higher-order terms}
In order to overcome this issue, we resort to another approach.
As can be checked from Eq.~(\ref{eq:appenInt0Kt}), the term in $(g-g_{12})n$ contributes to the expansion
only in terms of order $1/\sqrt{T}$ and more. We can thus neglect it here.
With this approximation, we have $\Eof \simeq \epsilon_{\vk} + \hbar\Omega_0$, so that we can simply rewrite Eq.~(\ref{eq:appenFluct}) in the form 
\begin{eqnarray}
G_\theta(0) & = &  \frac{1}{\pi}\sqrt{\frac{2\kB T}{\hbar^{2}n^2/2m}} \int_0^\infty du\
\big[ \textrm{coth}\left(u^2+\eta \right)-1 \big], \nonumber\\
\label{eq:appenFluct2}
\end{eqnarray}
where we defined the small parameter $\eta=\hbar\Omega_0/2\kB T$. We now split the integral into two parts.
For $u\ll\sqrt{\eta}$, $u^2+\eta \ll 1$ so that we can use the previous expansion of the coth function, and obtain 
\begin{eqnarray}
 \frac{1}{\pi}\sqrt{\frac{2\kB T}{\hbar^{2}n^2/2m}} 
 \int_0^{\sqrt{\eta}} du \left(\dfrac{1}{u^2+\eta}+\frac{u^2+\eta}{3}+...-1 \right)  \nonumber\\
\label{eq:appenFluct20uT}
\end{eqnarray}
Each term can then be exactly integrated. The first term gives a contribution linear in temperature,
which reads $\frac{m\kB T}{n\hbar^{2}L_{\theta}^{-1}}$.
Notice that, comparing to Eq.~(\ref{eq:appenExpansionPartielle}),
it yields only one half of the leading-order term linear in $T$.
All the other terms are of orders $1$, $1/T$, $1/T^2$,...,
thus strictly smaller than the $\sqrt{T}$ term we are looking for.
For $u\gg\sqrt{\eta}$, we can use the expansion
$\coth\left(u^2+\eta \right)\approx \coth(u^2)+\eta \coth^{(1)}(u^2)+...$,
where $\coth^{(n)}$ is the $n$-th derivative of $\coth$,
which yields 
\begin{eqnarray}
 \frac{1}{\pi}\sqrt{\frac{2\kB T}{\hbar^{2}n^2/2m}}
 \int_{\sqrt{\eta}}^{\infty}  du
 && \bigg\{ \left[\coth(u^2)-1\right]
  \nonumber\\
&& + \sum_{n\geq1} \dfrac{\eta^n}{n!} \coth^{(n)}(u^2)  \bigg\}.
\label{eq:appenFluct2uTinf}
\end{eqnarray}
Notice first that each term contains a contribution that is linear in $T$.
Indeed, their respective equivalents in 0 are non integrable and read $(\coth(u^2)-1)\sim_{u\rightarrow 0} 1/u^2$ and $\coth^{(n)}(u^2)\sim_{u\rightarrow 0} n!(-1)^n/u^{2n+2}$, so that all the terms in Eq.~(\ref{eq:appenFluct2uTinf}) scale once integrated as $1/\sqrt{\eta}$.
Together with the global prefactor $\sqrt{2\kB T}$, it yields a linear scaling.
The latter can be explicitly calculated by integrating the previous equivalents,
which yields
$\frac{4m\kB T}{n\pi\hbar^{2}L_{\theta}^{-1}}\times(1-1/3+1/5-1/7+...)
=
\frac{m\kB T}{n\hbar^{2}L_{\theta}^{-1}}$,
that is one half of Eq.~(\ref{eq:appenExpansionPartielle}).
Together with the contribution of the first part of the integral,
we thus recover exactly the same linear term as in the above section.
Then, coming back to Eq.~(\ref{eq:appenFluct2uTinf}), we can find the next order terms
by subtracting from each term its equivalent in $u=0$. 
The first correction reads
$\frac{1}{\pi}\sqrt{\frac{2\kB T}{\hbar^{2}n^2/2m}}  \int_{\sqrt{\eta}}^{\infty}  du~[\coth(u^2)-1-1/u^2]$.
The latter scales as $\sqrt{T}$ when $\eta\rightarrow 0$ since the function $u\rightarrow \coth(u^2)-1-1/u^2$
is integrable. One can then check that the contributions of the other terms will respectively scale as $1/\sqrt{T}$, $1/T^{3/2}$,...
We hence find the final expansion
\begin{equation}
G_\theta(\rr=0)\simeq
\frac{2m\kB T}{n\hbar^{2}L_{\theta}^{-1}}-\frac{I_1}{\pi}\sqrt{\frac{2\kB T}{\hbar^{2}n^2/2m}}
+ \textrm{O}(1),
\label{eq:appenExpansion}
\end{equation}
where $I_1=\int_0^\infty du\ [1/u^2-\coth(u^2)-1 ] \simeq 1.82$.

\end{appendix}


%

\end{document}